\pdfoutput=1
\documentclass[aps,prd,nofootinbib,twocolumn,citeautoscript,showkeys,superscriptaddress]{revtex4-1} 
\usepackage{amsmath}   
\usepackage{graphicx}  
\usepackage{dcolumn}   
\usepackage{bm}        
\usepackage{amssymb}   
\usepackage{epsfig}
\usepackage{epstopdf}  
\usepackage{subfigure}
\usepackage[utf8]{inputenc}
\usepackage{mdframed}
\usepackage{amsmath}
\usepackage{wasysym}
\usepackage[bookmarks=false]{hyperref} 
\hypersetup{pdfstartview=FitH,pdfhighlight=/O,colorlinks=true}
\usepackage{tabu}

\usepackage[toc]{appendix}
\usepackage{color,soul} 
\usepackage{datetime}

\DeclareMathOperator{\csch}{csch}

\newcommand{\bea}{\begin{eqnarray}}
\newcommand{\eea}{\end{eqnarray}}
\newcommand{\ba}{\begin{eqnarray}}
\newcommand{\ea}{\end{eqnarray}}

\newcommand{\beq}{\begin{equation}}
\newcommand{\eeq}{\end{equation}}
\newcommand{\beqa}{\begin{eqnarray}}
\newcommand{\eeqa}{\end{eqnarray}}
\newcommand{\beqar}{\begin{eqnarray*}}
\newcommand{\eeqar}{\end{eqnarray*}}

\newcommand{\eg}{{\it e.g.,}\ }

\newcommand{\PPsi}{\hat{\Psi}}
\newcommand{\hS}{\hat{S}}
\newcommand{\hG}{\hat{G}}

\newcommand{\req}[1]{(\ref{#1})} 

\begin{document}

\title{Echoes of Kerr-like wormholes} 
\author{Pablo Bueno} 
\email{pablo@itf.fys.kuleuven.be}
\affiliation{Institute for Theoretical Physics, KU Leuven,
Celestijnenlaan 200D, B-3001 Leuven, Belgium}
\author{Pablo A. Cano}
\email{pablo.cano@uam.es}
\affiliation{Institute for Theoretical Physics, KU Leuven,
Celestijnenlaan 200D, B-3001 Leuven, Belgium}
\affiliation{Instituto de F\'isica Te\'orica UAM/CSIC,
C/ Nicol\'as Cabrera, 13-15, C.U. Cantoblanco, 28049 Madrid, Spain}
\affiliation{Perimeter Institute for Theoretical Physics, Waterloo, ON N2L 2Y5, Canada}
	\author{Frederik Goelen}
\email{frederik.goelen@kuleuven.be}
\affiliation{Institute for Theoretical Physics, KU Leuven,
Celestijnenlaan 200D, B-3001 Leuven, Belgium}
\author{Thomas Hertog}
\email{thomas.hertog@kuleuven.be}
\affiliation{Institute for Theoretical Physics, KU Leuven,
Celestijnenlaan 200D, B-3001 Leuven, Belgium}
\author{Bert Vercnocke}
\email{bert.vercnocke@kuleuven.be}
\affiliation{Institute for Theoretical Physics, KU Leuven,
Celestijnenlaan 200D, B-3001 Leuven, Belgium}

\date{\today}

\begin{abstract}
Structure at the horizon scale of black holes would give rise to echoes of the gravitational wave signal associated with the post-merger ringdown phase in binary coalescences. We study the waveform of echoes in static and stationary, traversable wormholes in which perturbations are governed by a symmetric effective potential. We argue that echoes are dominated by the wormhole quasinormal frequency nearest to the fundamental black hole frequency that controls the primary signal.
We put forward an accurate method to construct the echoes waveform(s) from the primary signal and the quasinormal frequencies of the wormhole, which we characterize. 
We illustrate this in the static Damour-Solodukhin wormhole and in a new, rotating generalization that approximates a Kerr black hole outside the throat. Rotation gives rise to a potential with an intermediate plateau region that breaks the degeneracy of the quasinormal frequencies. Rotation also leads to late-time instabilities which, however, fade away for small angular momentum.

\end{abstract}

\maketitle
\section{Introduction}



The direct observation of gravitational waves produced during the merger of heavy compact objects \cite{Abbott:2016blz} offers exciting new opportunities for the study of black holes. Even though current observations do not yet probe the detailed structure of spacetime inside the light ring, one expects the near-horizon, strong gravity regime will gradually come into sight with future LIGO and Virgo observations and certainly with the next generation of gravitational wave observatories. 

This means that future gravitational wave observations have the potential to shed light on the nature of black hole (event) horizons and whether the near-horizon region exhibits any unexpected structure. Some would even argue there is a certain theoretical motivation for deviations of general relativity on horizon scales, because black hole event horizons are notoriously inconsistent with the basic principles of quantum mechanics as they are usually understood \cite{Lunin:2001jy,Almheiri:2012rt,Marolf:2017jkr}. In fact this has been a strong motivation for the program to construct horizonless alternatives to black holes, commonly referred to as exotic compact objects (ECOs) (see e.g., \cite{Mazur:2001fv,Schunck:2003kk,Morris:1988tu,Damour:2007ap,Mathur:2005zp,Holdom:2016nek}). Giving up the horizon, however, comes with a price which can include, depending on the kind of solution, various stability issues, the need for unphysical matter or even the lack of reasonable alternative collapse-processes or dynamics more generally, to name a few. In spite of these formidable drawbacks, the imminent possibility of testing the nature of (astrophysical) black holes demands a rigorous phenomenological study of alternatives.

It has been argued that the existence of any kind of structure at horizon scales would give rise to a series of ``echoes'' of the primary gravitational wave signal produced during the ringdown phase of a black hole merger \cite{Cardoso:2016rao,Cardoso:2016oxy}\footnote{See \cite{Barcelo:2010vc,Barcelo:2014cla} for a previous observation of the same phenomenon for electromagnetic signals.}. In fact, the LIGO data have already been analyzed on the presence of echoes \cite{Abedi:2016hgu,Ashton:2016xff,Abedi:2017isz}.

In order to use the full scientific potential of an observation of echoes to constrain ECO models, a solid theoretical understanding of the echo waveform is needed\footnote{See e.g., \cite{Price:2017cjr,Nakano:2017fvh,Volkel:2017kfj,Maselli:2017tfq} for recent work on this, and also \cite{Mark:2017dnq} which uses Green's function techniques to relate the train of echoes of the ECO from the ringdown response of the black hole.}. To make progress we put forward a method to model the waveforms of echoes in terms of the primary `black hole' signal and the quasinormal modes in two classes of traversable wormholes. The wormholes can be viewed as toy models for a class of ECOs in which perturbations are governed by a general, symmetric, one-dimensional effective potential. In particular the potentials exhibit besides the usual bump characteristic of black holes a second bump separated from the first one by a `distance' $L$, as illustrated in Fig. \ref{fig1}. 

The ringdown phase is governed by the quasinormal modes (QNMs) of the final object. In Section \ref{QNMs} we therefore first compute the quasinormal frequencies (QNFs) in ECO spacetimes with symmetric potentials. We show the QNFs can be obtained from the reflection coefficients associated with the first bump of the potential only. We find that the real parts of the QNFs are, up to a constant, approximately given by $n \pi/L$, even for very high QNFs. We also find that the imaginary parts are generically $\leq \mathcal{O}(L^{-3})$ and therefore much smaller than their black hole counterparts. This in turn means that the damping of the signal is limited. Instead the waveform becomes approximately periodic, with period $\sim 2L$, after the first echo, with minor deformations induced by the failure of the consecutive real parts of the QNFs to differ exactly by $\pi/L$. 

In Section \ref{ECHOE} we exploit these features to construct the full waveforms of the subsequent echoes from the first echo waveform. The shape of the latter can be approximately determined from the frequency content of the primary signal. In particular, we point out that the most relevant wormhole QNF is the one nearest to the fundamental black hole QNF, which dominates the primary signal.
This yields a general procedure to determine the waveforms of echoes that is an alternative to that of \cite{Mark:2017dnq}, which uses the reflection and transmission coefficients of the single-bump potential together with the near horizon and asymptotic black hole waveforms. In section \ref{markos}, we also generalize the method in \cite{Mark:2017dnq} to rotating solutions.

In Section \ref{DSW} we apply our method to construct the echo waveforms in the time domain in the Schwarzschild-like wormhole of Damour and Solodukhin \cite{Damour:2007ap}, starting from a primary Gaussian waveform. We also compare our results for the QNFs with the ones obtained by replacing the potential with the double P\"oschl-Teller potential.

Finally in Section \ref{KW} we apply our method in an ECO modification of the Kerr black hole which turns it into a wormhole that reproduces Kerr's spacetime away from the throat. The behavior of scalar perturbations can again be described in terms of a one-dimensional, symmetric `double-bump' potential with an intermediate plateau region with height proportional to the angular momentum of the wormhole. We reproduce the features of the QNFs mentioned earlier, and observe that the rotation leads to a Zeeman-like breaking of the degeneracy of the QNFs, which increases with angular momentum. We also identify the presence of unstable QNMs and comment on their impact on the angular momentum of the final object. Finally, we use our method to reconstruct a series of echo-waveforms, by modelling the leading one as a Gaussian controlled by the fundamental QNF of the corresponding Kerr black hole. 

We conclude in Section \ref{conclusions}, and include an Appendix \ref{sdp} with a number of examples of potentials for which part of the analysis of Section \ref{DP} can be done analytically.

\section{Scattering in symmetric potentials}\label{DP}

\begin{figure}[t!]
	\centering 
	\includegraphics[width=8.9cm]{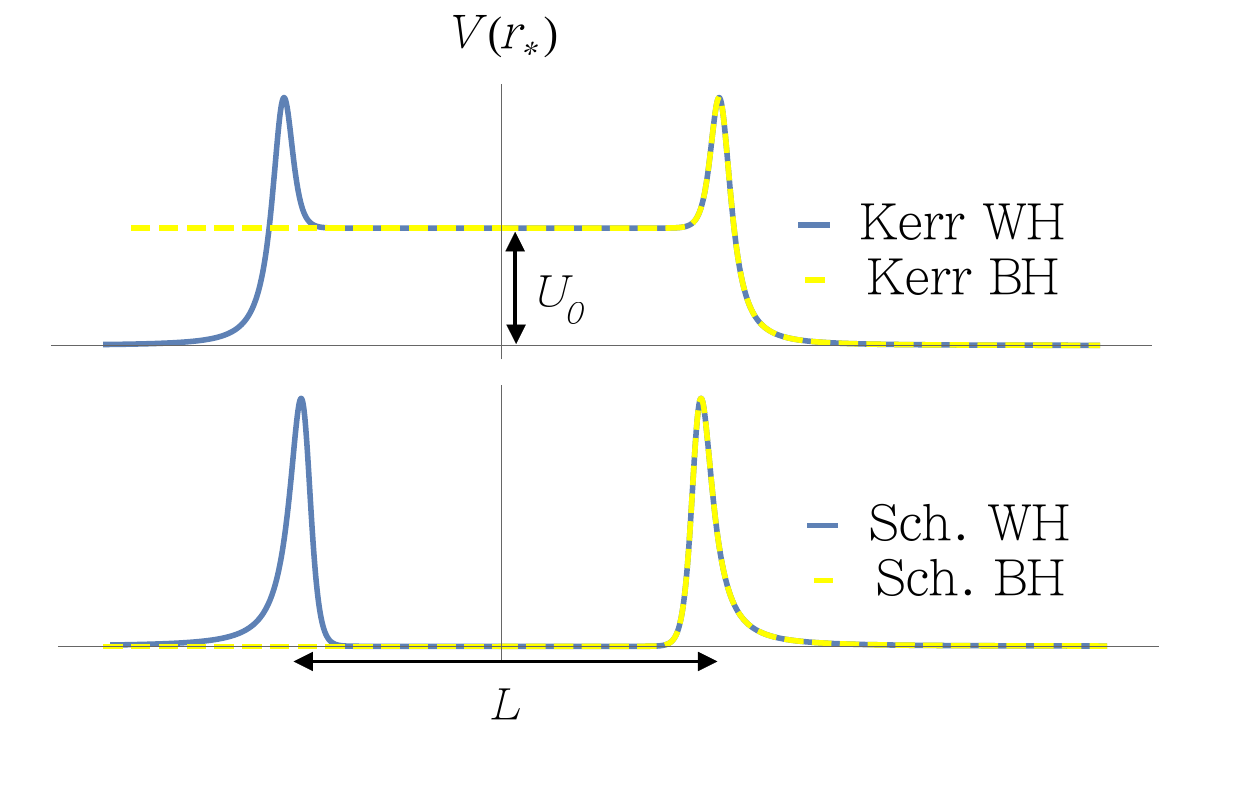}
	\caption{Effective potentials for a probe scalar field mode in a Kerr-like and Schwarzschild-like wormhole (blue) and black hole (dashed) background as a function of the tortoise coordinate $r_{*}$. The Kerr black hole potential tends to a frequency-dependent constant, $V(r_*)\rightarrow U_0(\omega)\equiv \omega_0(2\omega-\omega_0)$ as $r_{*}\rightarrow -\infty$ with $\omega_0= a m/(2 M (M+\sqrt{M^2-a^2})) $, where $a M$ and $M$ are the angular momentum and mass of the black hole, respectively, and $m$ is the ``magnetic quantum number'' of the mode. The plots above correspond to $l=m=2$, but the qualitative features of the potentials are the same for general $(l,m)$. }
	\label{fig1}
\end{figure}

In this section, we study various general aspects of scattering in one-dimensional double potentials that consist of two mirror-symmetric copies of single-bump potentials glued at $x=0$ and decaying to zero as $x \rightarrow \pm \infty$, possibly with an intermediate plateau region (cf. Fig.\ \ref{fig1}). Potentials of this kind capture the effective potentials felt by scalar perturbations of Schwarzschild-like and Kerr-like wormhole spacetimes as we discuss below.

\subsection{Quasinormal modes and frequencies}\label{QNMs}
The problem of characterizing the quasinormal modes (QNMs) and frequencies of scalar perturbations on traversable-wormhole backgrounds  can be reduced to a one-dimensional scattering problem governed by a wave equation of the following form\footnote{\label{fofy}
We denote functions in the frequency domain with a hat, e.g., $\hat \Psi(x,\omega)$. Solutions in the time domain are then given by $\Psi(x,t)= \frac 1 {2\pi}\int_{-\infty }^{+\infty} d\omega \PPsi (x,\omega)$, and at late times by a sum of quasinormal modes.}
\begin{equation}
\frac{d^2\PPsi}{dx^2}+(\omega^2-V(x,\omega))\PPsi=0\, ,
\label{QNMproblem}
\end{equation}
with boundary conditions
\begin{equation}
\lim_{x\rightarrow\pm\infty}\PPsi(x)\sim e^{\pm i\omega x}\, .
\end{equation}
and with a potential $V(x)$ that approximately takes the form 
 \begin{equation}
 V(x,\omega)=\theta(x)U(x-L/2,\omega)+(x \leftrightarrow -x)\, .
 \end{equation}
Here, $U(x,\omega)$ is the single-bump potential in the black hole background that matches the wormhole spacetime outside the throat, $\theta(x)$ is the Heaviside step function and $L$ represents the separation between both potential bumps in the wormhole background. The potential $U$ has a maximum $U_{\rm max}$ and a characteristic width $U_{\rm width}$. To model the wormhole potential $V$ we approximate $U(x)=U_0$ for all $x<x_0$ with $|x_0|<L/2$. 
 
For the static Schwarzschild-like wormholes we consider in Section \ref{DSW}, $U_0=0$ for general perturbations. By contrast, for the scalar perturbations on the Kerr-like wormhole we consider in Section \ref{KW}, we will find (cf. \eqref{omeg}) that $U_0$ is non-vanishing and frequency-dependent, and such that $\sqrt{\omega^2-U_0(\omega)}=\omega-\omega_0$, with $\omega_0$ a real constant. 

To determine the QNFs, we start by solving the following scattering problem 
\begin{equation}\label{pu}
\PPsi_U(x)=\begin{cases}
A' e^{i \omega x}+B'e^{-i\omega x}  \quad  &{\rm if}\quad x\rightarrow+ \infty\, ,\\
Ae^{i (\omega - \omega_0)  x}+Be^{-i (\omega - \omega_0)  x} \quad  &{\rm if}\quad x\rightarrow-\infty\, ,
\end{cases}
\end{equation}
for the potential $U(x)$, to determine the transfer matrix $\mathrm{T}$ such that
\begin{equation}
\begin{pmatrix}
A'\\
B'
\end{pmatrix}=
\mathrm{T}
\begin{pmatrix}
A\\
B
\end{pmatrix}\, .
\end{equation}
Using this, we can construct the transfer matrix of the full double potential $V(x)$, which we denote $\mathbb{T}$. Note that the transfer matrix of $U(-x)$ is given by $\sigma \mathrm{T}^{-1}\sigma$, where $\sigma=\begin{pmatrix}
0&1\\
1&0
\end{pmatrix}$. Then, we have
\begin{equation}
\mathbb{T}=\mathrm{T}\begin{pmatrix}
e^{i  (\omega - \omega_0)   L}&0\\
0&e^{-i  (\omega - \omega_0)   L}
\end{pmatrix}\sigma \mathrm{T}^{-1}\sigma\, .
\end{equation}
QNMs are characterized by the condition that there are no incoming waves, i.e., $B'=0$, $A=0$. This implies $\mathbb{T}_{22}=0$, which can be written in the suggestive form\footnote{An equivalent expression valid for more general ECOs was obtained in \cite{Mark:2017dnq} for $\omega_0=0$.}
\begin{equation}\label{refle}
 e^{-i(\omega_n - \omega_0) L}= -e^{i\pi n} R_{\rm BH}(\omega_n)\, ,
\end{equation}
where $-e^{i \pi n} =(-1)^{n+1}$ is the parity and $R_{\rm BH}(\omega) \equiv - \mathrm{T}_{21} / \mathrm{T}_{22}$
is the reflection coefficient of $U(x)$ for a wave coming in from $-\infty$. Eq.\ \req{refle} enables one to find the QNFs of the double potential $V(x)$ once $R_{\rm BH}(\omega)$ is known. We will use this below to compute the QNFs of Schwarzschild-like and Kerr-like wormholes.
  
For a given QNF, the corresponding QNM is given by
 \begin{equation}
 \PPsi(x)=\theta(x)\PPsi_U(x-L/2)- e^{-i\pi n} (x \leftrightarrow -x)\, ,\label{eq:psi_from_psi_U}
 \end{equation}
 where $\PPsi_U(x)$ is the solution to the scattering problem \req{pu} in the black-hole potential with $B'= 0$.
The function $\PPsi(x)$ constructed this way will turn out to be an excellent approximation to the exact mode solution in wormhole backgrounds, which solves \req{QNMproblem}, since $|V(-L/2)-U_0|\ll1$. 

We note that $\omega=\omega_0$ is always a solution of \req{refle}, given that $R_{\rm BH}(\omega_0)=-1$.
This solution is non-physical since it would mean there is no wave inside the cavity, but it allows us to obtain approximate solutions for $\omega \sim \omega_0$. In particular, if $L$ is much greater than the width $U_{\rm width}$ then the first few QNFs will be of order $\omega_n-\omega_0\sim 1/L$. We can obtain approximate solutions of \req{refle} for these modes $\omega_n$ by Taylor expanding $R_{\rm BH}$ around $\omega_0$,
\begin{equation}\label{rtay} 
R_{\rm BH}(\omega_n)=-1+\sum_{k=1}^{\infty}\frac{R_{\rm BH}^{(k)}(\omega_0)}{k!}(\omega_n-\omega_0)^k\, .
\end{equation}
and similarly writing $\omega=\sum_{k=0}^{\infty}c_k/L^k$. This yields, up to second order, 
\begin{align}\label{ttt}
 &\omega_n \approx \omega_0+\frac{\pi n}{L}\Big\{1- \frac{iR_{\rm BH}'(\omega_0)}{L}  \\ \notag & -\frac{i}{2L^2}\Big[n\pi \left( R_{\rm BH}'(\omega_0)^2+R_{\rm BH}''(\omega_0) \right)-2iR_{\rm BH}'(\omega_0)^2\Big]
\Big\}
 \, ,
 \end{align} 
 where the corrections are of order $\mathcal{O}\left(\vert  U_{\rm width }/L \vert \right)^3$. However, this does not mean that this is an accurate expression for the real and imaginary parts of the QNFs separately. Usually, it is the case that ${\rm Re}\, \omega_n \gg {\rm Im}\, \omega_n$. Thus, the result above gives a good approximation of the real part of $\omega_n$ only. To get an approximate solution for the imaginary part, we take the absolute value of \req{refle}, leading to
\begin{equation}
{\rm Im}\,\omega_n=\frac{1}{L}\log|R_{\rm BH}(\omega_n)|\simeq\frac{|R_{\rm BH}(\omega_n)|-1}{L}\, ,
\end{equation}
where we used that $|R_{\rm BH}(\omega_n)|\approx 1$. Substituting \req{ttt} then yields an accurate approximation to ${\rm Im}\,\omega_n$. In sum, we get
\begin{align}\label{ttto}
& {\rm Re} \,\omega_n\simeq \omega_0+\frac{\pi n}{L}\left(1+ \frac{ {\rm Im} \,R_{\rm BH}'(\omega_0)}{L} \right)\, , \\  \label{tttt}
&{\rm Im} \,\omega_n\simeq \frac{|R_{\rm BH}({\rm Re} \,\omega_n)|-1}{L}\, .
\end{align}
Therefore, up to an overall shift, the real parts are all approximately proportional to $\pi/L$. 
The imaginary parts are $\mathcal{O}(L^{-3})$ and hence much smaller. In fact, by taking the absolute value in \eqref{rtay} and assuming $|R_{\rm BH}|\leq 1$ it follows that $ {\rm Re} \,R'_{\rm BH}(\omega_0)=0$, which implies that\footnote{For the rotating case, it is not always true that $|R_{\rm BH}|\leq 1$ and thus some of the imaginary parts can scale with $1/L^2$, as we mention in \ref{QNFKL}.} ${\rm Im} \,\omega_n\sim \mathcal{O}(L^{-3})$. More precisely, this argument shows that ${\rm Im}(\omega_n)\le \mathcal{O}(L^{-3})$. Indeed for the Schwarzschild black hole potential $U$ (see eq. \req{SchwPot} with $\lambda=0$) numerical evaluation of \req{ttt} suggests that the $\mathcal{O}(L^{-3})$ terms are only non-vanishing for $l=0$, where $l$ is the angular number of the perturbation. This means that for any $l>0$ we have ${\rm Im}(\omega_n)=o(L^{-3})$, consistent with \cite{Cardoso:2017njb}, where it was argued that the imaginary parts generically scale with $\sim L^{-(2l+3)}$.

Let us now turn to the high-frequency regime. For that, we momentarily set $\omega_0=0$. In this regime we can regard the potential $U$ in \eqref{QNMproblem} as a perturbation.
In particular, writing $\PPsi(x)=A(x) e^{-i \omega x}$ 
and expanding the prefactor $A(x)$ in powers of $U$, $A=A_0+A_1 +A_2 +\dots$, where $A_n\sim \mathcal{O}(U^n)$, we can solve \eqref{QNMproblem} order by order in $U$. The leading terms are given by
\begin{equation}\label{pio}
\frac{d^2A_0}{dx^2}-2i\omega \frac{dA_0}{dx}=0\, ,\quad \frac{d^2A_1}{dx^2}-2i\omega \frac{dA_1}{dx}=U A_0\, .
\end{equation}
The boundary condition \eqref{pu} implies $A_0=e^{2i\omega x}$. Further, since the width of the potential, which sets the scale on which $A_1$ changes, is much larger than $\omega^{-1}$, we have $\left|\omega \frac{dA_1}{dx}\right|\gg \left|\frac{d^2A_1}{dx^2}\right|$.  Hence, an approximate solution is given by
\begin{equation}
A_1(x)=\frac{1}{2i\omega}\int_{x}^{+\infty}dx'e^{2i\omega x'}U(x')\, .
\end{equation}
Plugging this back in \req{pio}, we observe that the second equation fails to be satisfied by $\frac{1}{\omega} \frac{d \log U(x)}{dx}$, which is indeed much smaller than $1$ for any reasonable bump-like potential in the high-frequency limit.

Then, we find
\begin{equation}
\PPsi(x)=e^{i\omega x}+e^{-i\omega x} A_1(x)\, ,
\end{equation}
which satisfies  $\PPsi(x)\rightarrow e^{i\omega x}$ when $x\rightarrow\infty$ and $\PPsi(x)\rightarrow e^{i\omega x}+e^{-i\omega x} A_1(-\infty)$ as $x\rightarrow-\infty$. Thus the reflection coefficient is nothing but $A_1(-\infty)$:
\begin{equation}
R_{\rm BH}(\omega)=\frac{1}{2i\omega}\int_{-\infty}^{+\infty}dx'e^{2i\omega x'}U(x')\, \,\, \,  {\rm for} \, \, \, \,\omega\rightarrow\infty\, .
\end{equation} 
This is valid as long as the integral is convergent. 

As an illustration we evaluate this explicitly for the P\"oschl-Teller potential $U(x)=V_0 \operatorname{sech}^2(\alpha x)$. This gives (see appendix \ref{sdp}) 
\begin{equation}
R_{\rm BH}(\omega)=\frac{\pi V_0}{i\alpha^2\sinh (\pi \omega/\alpha)}\simeq \frac{2\pi V_0}{i\alpha^2}e^{-\pi\omega/\alpha}\, ,
\end{equation}
since $\omega/\alpha \gg1$ for high frequencies. Using Eq.\ \req{refle} we then obtain, for large $n$,
\begin{equation}
\omega_n\simeq \frac{\pi n}{L+i \pi/\alpha}\, ,
\end{equation}
which is the asymptotic behavior in the sense that $\lim_{n\rightarrow\infty}\frac{(L+i\pi/\alpha)\omega_n}{n\pi}=1$. Hence, 
\begin{equation}\label{omre}
{\rm Re}(\omega_{n+1}-\omega_n)=\frac{\pi L}{L^2+\pi^2/\alpha^2}\simeq \frac{\pi}{L}\, ,
\end{equation}
where we used $L\gg 1/\alpha$. 

Hence, the spacing of the real parts of consecutive QNFs is approximately $\pi/L$ for general $n$. Below we show numerically this remains true for general double-bump potentials. The underlying reason is that the rate of variation of the reflection coefficient is controlled by the width of the potential, which we assume to be much smaller than $L$\footnote{In particular, a shift of $\omega\rightarrow \omega+\pi/L$ in the frequency barely changes $R_{\rm BH}$, but it induces a minus sign in the left-hand side of \req{refle}, meaning that if $\omega$ is a solution, then $\omega+\pi/L$ is close to another solution with opposite parity.}.
This also means that the QNF spectrum of wormholes contains frequencies with an arbitrarily large real part, in contrast to the QNF spectrum of black holes.

\subsection{Time dependence and echoes}\label{ECHOE}

We have seen that the spectrum of QNFs in wormhole backgrounds differs drastically from that of black holes. Here we use the spectral features of the QNMs to derive the presence of echoes in characteristic time-domain signals.

Consider a primary signal produced by a `black hole' perturbation near one of the maxima of the potential.  After relaxation, this is given by a linear combination of QNMs which, at a given spatial point, amounts to a sum of quasi-harmonic terms,
\begin{equation}
\Psi(t)=\sum_{n=-\infty}^{\infty}c_{n}e^{-i\omega_{n} t}\, ,
\label{TSs}
\end{equation}
for some coefficients $c_n$.

In wormhole backgrounds with a large separation between both maxima, primary signals of this kind are not influenced by the second bump. They are similar to signals produced by similar perturbations of a black hole, and naturally expressed as a superposition of black hole QNMs. At a later stage, however, the perturbation is reflected off the second bump. This produces a signal which one expects, regardless its primary source, consists of a sum of wormhole QNMs, 
\begin{equation}\label{jk}
\Psi(t)\approx e^{-i \omega_0t} \left(\sum_{n=-\infty}^{\infty}c_{n}e^{-i n\pi t/L}e^{{\rm Im}\,\omega_{n}t}\right) \, .
\end{equation}
The factor in between brackets only slowly decays and is approximately periodic, with period $T=2L$, since ${\rm Im}\,\omega_{n}\ll {\rm Re} \, \omega_{n} $ (cf. \eqref{tttt})\footnote{The factor $e^{-i\omega_0 t}$ produces an additional effect, but does not affect the amplitude, so modulated waves are still modulated.}.
Hence, \req{jk} approximately takes the form of a Fourier series with period $T$. This means that on longer timescales, the primary signal is approximately repeated periodically, giving rise to the echoes. 


This observation also suggests that the coefficients of subsequent echoes can be obtained from the waveform of the first echo $\Psi_{\rm {1^{st}} echo}^{(0)}(t)$. In particular we get,
\begin{equation}\label{coeff2}
c_{n}=\frac{1}{2L}\int_0^{2L}dt \Psi_{\rm {1^{st}} echo}^{(0)}(t) e^{i\omega_{n}t}\, .
\end{equation}
It is a universal feature of wormholes that once the first echo is identified, the subsequent waveform can be accurately constructed from \req{TSs} using the coefficients \req{coeff2}. We illustrate this in the next section.  


It remains to identify the waveform of the leading echo. On general grounds one expects this to be dominated by the wormhole QNF that is nearest to the lowest black hole QNF, which dominates the primary signal. This is borne out by the examples below. Hence it is natural to model the first echo as a Gaussian wave packet of the form
\begin{equation}
\Psi_{\rm {1^{st}} echo}^{(0)}(t)=e^{-i\omega^{\rm BH}_{\rm 0}(t-t_0)}e^{-\frac{(t-t_0)^2}{2\tau ^2}}\, ,
\label{WP}
\end{equation}
where $\omega^{\rm BH}_{0}$ is the leading QNF of the black hole.
Assuming that $\tau \ll L$ we can then use the method above to determine the coefficients $c_{n}$ of the subsequent echoes, yielding
approximately
\begin{equation}\label{WPC}
c_{n}=\sqrt{\frac{\pi}{2}}\frac{\tau }{L}\exp{\left[i\omega_{n}t_0-\frac{\tau ^2}{2}(\omega_{n}-\omega_0^{\rm BH})^2\right]}\, .
\end{equation}
Substituting these in \req{TSs}, we obtain an echo waveform which should resemble the actual waveform produced by a Gaussian primary perturbation. We use this construction to model the echoes waveform corresponding to a rotating Kerr-like wormhole  in Section \ref{KW}.

\subsection{Reconstruct waveform from black hole result}\label{markos}


A more sophisticated method to construct the waveform of the first echo using black hole information was recently developed in \cite{Mark:2017dnq}. This uses the asymptotic and near-horizon signals produced by a single-bump black hole, as well as its reflection and transmission coefficients, to construct the asymptotic signal produced by a static wormhole. Here we generalize this to rotating wormholes.

Stated accurately, we show how to obtain the asymptotic solution $\PPsi(x\to\infty)$ for the wormhole background from the wave function of the black hole scattering problem with the potential $U$.
Eq.\ \req{eq:psi_from_psi_U} shows how to obtain $\PPsi(x,\omega)$ from the solution $\PPsi_U$ to the scattering problem with a single-bump potential $U$ and source $\hS$:
\begin{equation}
\frac{d^2 \PPsi_U}{dx^2} + (\omega^2 - U(x)) \PPsi_U = \hS(x,\omega)\,.\label{eq:diffU}
\end{equation}
While $\PPsi$ obeys the outgoing boundary conditions $ \lim_{x\rightarrow\pm\infty}\PPsi(x)\sim e^{\pm i\omega x}$, the wave function $\PPsi_U$ itself only has outgoing boundary conditions as $x\to \infty$, i.e., as in eq.\ \req{pu} with $B'=0$, and $B/A = R_{\rm BH}(\omega)e^{	i (\omega - \omega_0) L}$. This means that $\PPsi_U$ is \emph{not} the physical solution of the wave function for the black hole background. The solution $\PPsi_{\rm BH}$ for scattering in the Kerr black hole solves the same differential equation \eqref{eq:diffU}, but with boundary conditions
\begin{equation}
\PPsi_{\rm BH}(x)=\begin{cases}
e^{i\omega x}  \quad  &{\rm if}\quad x\rightarrow+ \infty\, ,\\
e^{-i (\omega - \omega_0)  x}  \quad  &{\rm if}\quad x\rightarrow-\infty\,.
\end{cases}
\end{equation}

The remaining issue is then to obtain $\PPsi_U$ from $\PPsi_{\rm BH}$. 
First, express the wavefunction in terms of the  Green's function as $\PPsi= \int_{-\infty}^{+\infty} dx' \hS(x',\omega) \hG(x,x')$,
where $\hG(x,x')$ is the solution to 
\begin{equation}
\frac{d^2 \hat{G}}{dx^2} + (\omega^2 - U(x))\hat{G} = \delta(x,x')\,,\label{eq:deltasource}
\end{equation}
and consider Green's functions $\hG_U, \hG_{\rm BH}$ with the same boundary conditions as respectively $\PPsi_U,\PPsi_{\rm BH}$.
Standard results give the black hole Green's function as
\begin{equation}
\hG_{\rm BH}(x,x') =  \frac{\PPsi_+(\max(x,x'))\PPsi_-(\min(x,x'))}{W_{\rm BH}}\,,
\end{equation}
with the functions $\PPsi_{\pm}$ independent solutions of \eqref{eq:deltasource} for outgoing waves at infinity ($\PPsi_+$ with $B'_+=0$ in \eqref{pu}) and outgoing waves at the horizon ($\PPsi_-$ with $A_-=0$ in \eqref{pu}). The Wronskian is $W_{\rm BH}(\omega) = 2 i (\omega -\omega_0)B_+$.

A straightforward generalization of the static results of \cite{Mark:2017dnq} then gives 
\begin{equation}
\hG_U = \hG_{\rm BH}  + {\cal K}(\omega) \frac{\PPsi_+(x) \PPsi_+(x')}{W_{\rm BH}}\, ,
\end{equation}
with
\begin{equation}\label{trans}
{\cal K}(\omega) =   \frac{T_{\rm BH} R_{\rm BH} e^{ 2i (\omega-\omega_0) L}}{1 - R_{\rm BH}^2 e^{2i(\omega-\omega_0) L}}\,.
\end{equation}
One can integrate this over the source to get the full wavefunction. Note that the poles $\omega_n$ of $\cal K$ are precisely the QNFs of the wormhole, as expressed in eq.\ \eqref{refle}. Each term in the Green's function also has poles  at the QNFs of the black hole from the zeroes of the Wronskian $W_{\rm \rm BH}$. Those cancel in $\hG_U$ and hence $\PPsi_U$ only has QNFs $\omega_n$.

Observe now that the asymptotic limit of the wavefunction is given by $\lim_{x\to +\infty} \PPsi(x) = \lim _{x\to +\infty}\PPsi_U(x-L/2)$. We find it has a correction to the black hole waveform of the form (cf. footnote \ref{fofy}):
\begin{equation}
\PPsi(x) = \PPsi_{{\rm BH}}(x-L/2) + \PPsi_{\rm corr}(x-L/2)\,,\quad x\to \infty\, ,
\end{equation}
with $\PPsi_{\rm corr}$ given by the transfer function $\cal K$ and the near-horizon part of the black hole wavefunction,
\begin{align}\label{marktime}
\PPsi_{\rm corr}(x) =\,& {\cal K}(\omega)e^{i\omega x}\lim_{x\to -\infty} \left(\PPsi_{\rm BH}e^{i (\omega -\omega_0)x}\right)\, .
\end{align}
The factor in brackets is evaluated at the black hole horizon and represents the near-horizon response. It is multiplied by an exponential factor such that it has no overall $x$-dependence. Then the two terms in $\PPsi$ indeed have the correct asymptotic behaviour at large $x$.
This can be generalized to other ECOs. Given the mass and spin of a given black hole are known, all one needs is the transfer function $\cal K$ of the ECO one wishes to study.

\section{Damour-Solodukhin wormhole}\label{DSW}

We turn now to our first explicit example, the static Schwarzschild-like wormhole considered by Damour and Solodukhin (DS) in \cite{Damour:2007ap} with metric
\begin{equation}\label{wormhole}
ds^2=-(f(r)+\lambda^2) dt^2+\frac{dr^2}{f(r)}+r^2 d\Omega_{(2)}^2 \, , 
\end{equation}
where $f(r)=1-2M/r$. Naturally, when $\lambda=0$, this reduces to the usual Schwarzschild black hole metric. For non-zero values of the parameter $\lambda^2$, however, \req{wormhole} is no longer a solution of Einstein's equations and the manifold structure changes drastically. In particular, the Einstein tensor of \req{wormhole} has a vanishing time-time component, while $G_{rr},G_{\theta\theta},G_{\phi\phi}\sim \lambda^2 $. 
This means, among other things, that matter with vanishing energy density would be required to sustain such a gravitational configuration. However, we regard these wormholes here in the first place as tractable toy models to study some of the phenomenological implications of ECOs.

In \req{wormhole}, $t$ does not correspond to the time of an asymptotic observer. It is convenient to redefine $t\rightarrow t/\sqrt{1+\lambda^2}$ and $M\rightarrow M(1+\lambda^2)$, to get 
\begin{equation}\label{wormhole2}
ds^2=-f(r)dt^2+\frac{dr^2}{g(r)}+r^2 d\Omega_{(2)}^2\, ,
\end{equation}
where now
\begin{equation}
f(r)=1-\frac{2M}{r}\, , \quad  g(r)=1-\frac{2M(1+\lambda^2)}{r}\, .
\end{equation}

Let us now consider a massless test scalar field on the background metric \req{wormhole2}. After decomposition in spherical harmonics, the scalar perturbations are described by the equation
\begin{equation}
\left[\frac{\partial^2}{\partial t^2}-\frac{\partial^2}{\partial r_*^{2}}+V_{l}(r)\right]\Psi_{l}(t,r_*)=0\, ,
\label{timeeq}
\end{equation}
where
\begin{equation}\label{SchwPot}
V_{l}(r)=\frac{l(l+1)f(r)}{r^2}+\frac{(f(r)g(r))'}{2r}\, ,
\end{equation}
is the scalar Regge-Wheeler potential, and $r_*$ is the tortoise coordinate, defined implicitly by the relation $dr_*/dr=1/\sqrt{f(r)g(r)}$. If we further separate the time and radial coordinates as $\Psi_{l}(t,r)=e^{-i\omega t}\PPsi_l(r_*)$ we get the following equation for the radial wavefunction
\begin{equation}
\frac{d^2\PPsi_{l}}{dr_*^{2}}+\left(\omega^2-V_{l}(r)\right)\PPsi_{l}=0\, .
\end{equation}

The relation between $r$ and $r_*$ for the wormhole metric can be conveniently written in terms of an auxiliary variable $\rho$ as
\begin{eqnarray}
r/M&=&2+\lambda^2(1+ \cosh\rho)\, ,\\
r_{*}/M&=&(2+\lambda^2)\rho+\lambda^2\sinh\rho\, .
\end{eqnarray}
Letting $\rho$ range from $-\infty$ to $+\infty$, we observe that $r\ge 2M(1+\lambda)^2$, while $r_*$ takes all values in the real line. Indeed, $r_*$ and $-r_*$ correspond to the same $r$. Furthermore, $r_*=0$ corresponds to the throat position $r=2M(1+\lambda^2)$. The wormhole structure becomes evident from this perspective.

The relation $r_*(r)$ can be written explicitly as
\begin{equation}
\begin{aligned}
r_*&=\pm M\left( \sqrt{(r/M-2)(r/M-2(1+\lambda^2))}\right. \\ \notag &+(2+\lambda^2)\cosh^{-1}\left[\lambda^{-2}(r/M-2)-1\right] \Big)\, .
\end{aligned}
\end{equation}
Now, we are interested in the regime of small $\lambda^2$, for which the metric is essentially indistinguishable from a Schwarzschild black hole to an external observer. 
For $\lambda^2 \ll 1$, we get
\begin{equation}
r_*= r+2M\log\left(\frac{r}{2M}-1\right) +2M\left(-1+\log\left(4\lambda^{-2}\right)\right)\, ,
\end{equation}
up to $\mathcal{O}(\lambda^2)$ corrections. This can be written more suggestively as
\begin{equation}
r_*= r_*^{\rm BH} +\frac{L}{2}\, ,\,\, {\rm where} \, \, L\equiv4M\left(-1+\log\left(4\lambda^{-2}\right)\right)\, .
\end{equation}
Here, $L$ represents the length of the wormhole throat, and therefore it is approximately the distance between the maxima of the potential $V_l(r(r_*))$. In addition, when $\lambda\ll1$ we get $V_l(r)= V_l^{\rm BH}(r)+\mathcal{O}(\lambda^2)$. Altogether, we conclude that the potential $V_l(r_*)$ is very approximately
\begin{equation}
V_l(r_*)=\theta(r_*)V_l^{\rm BH}(r_*-L/2)+(r_*\leftrightarrow -r_*)\, ,
\label{StaticPot}
\end{equation}
where $V_l^{\rm BH}(\cdot)$ represents the black hole potential as a function of the black hole tortoise coordinate  $r_*^{\rm BH}$. Hence, the limit $\lambda\ll 1$ of this model essentially coincides with a double copy of the Schwarzschild black hole connected through a throat of length $L$. Note that the dependence of $L$ in $\lambda^2$ is logarithmic, so even if this is small --- say, $\lambda^2 \sim \ell_{\rm P}/M$ --- the effect will be measurable through the detection of perturbations which traverse the throat. This phenomenon is characteristic of general ECOs, as observed in \cite{Cardoso:2016oxy,Cardoso:2016rao}.

\subsection{Quasinormal modes and frequencies}
The effective potential $V_l(r_*)$ in \req{StaticPot} takes the form of those studied in the previous section, so we can apply the method put forward there. In order to compute the QNFs, we need to obtain the reflection coefficient of the Schwarzschild potential as a function of the frequency, $R_{\rm BH}(\omega)$. This can be easily obtained numerically by solving the problem in which there is only an outgoing wave for $x\rightarrow +\infty$. Alternatively, we can get an approximate solution by replacing the true wormhole potential by a double P\"oschl-Teller potential for which we know $R(\omega)$ analytically (cf. Appendix \ref{sdp}). The second alternative has the advantage that it allows us to construct approximate analytic expressions for the wormhole QNMs. 

\begin{figure}[ht!]
\centering 
	\includegraphics[width=8.5cm]{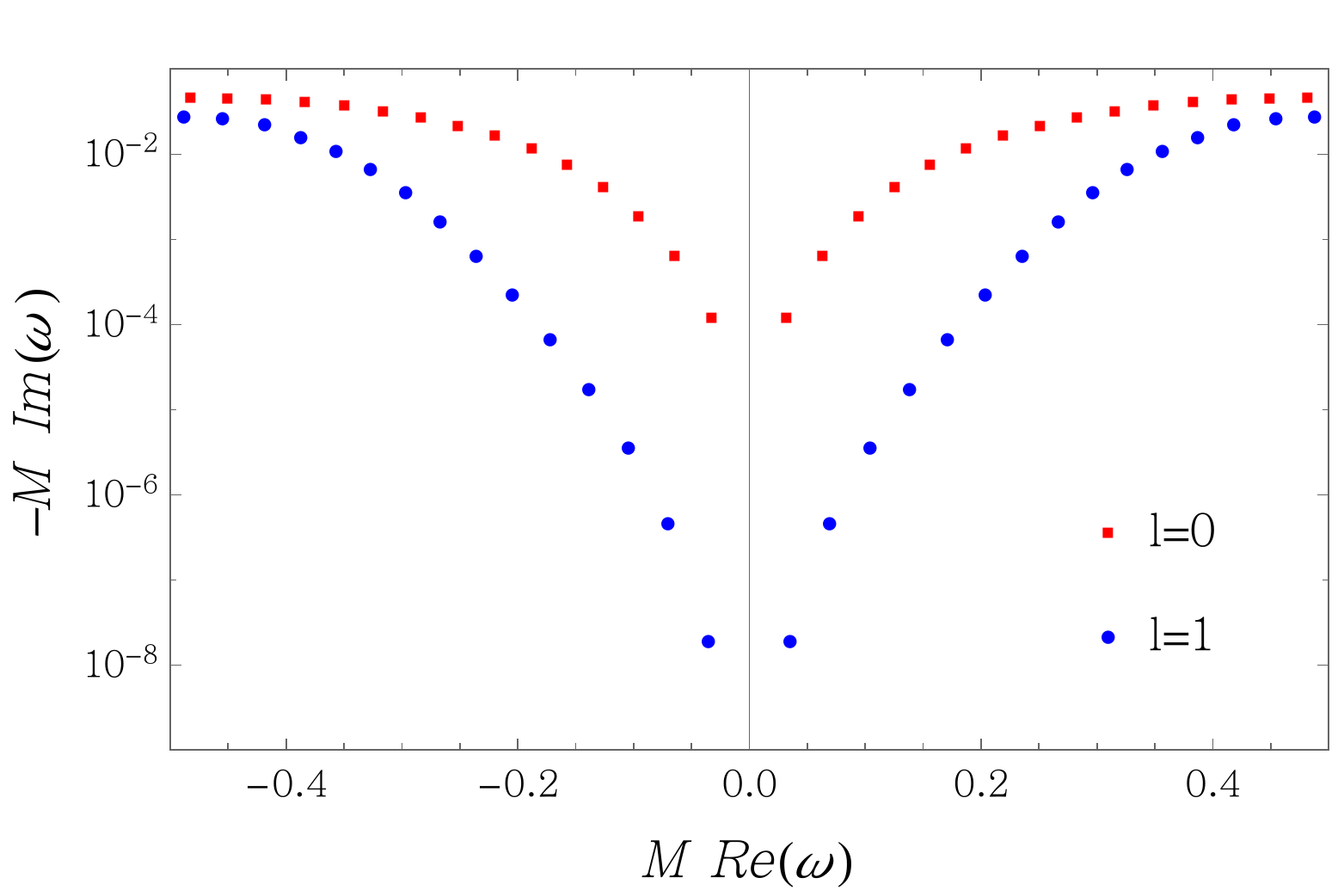}
	\caption{Quasinormal wormhole frequencies for $\lambda^2=10^{-10}$.}
	\label{QNFplot1}
\end{figure}

\begin{figure}[ht!]
\centering 
	\includegraphics[width=9cm]{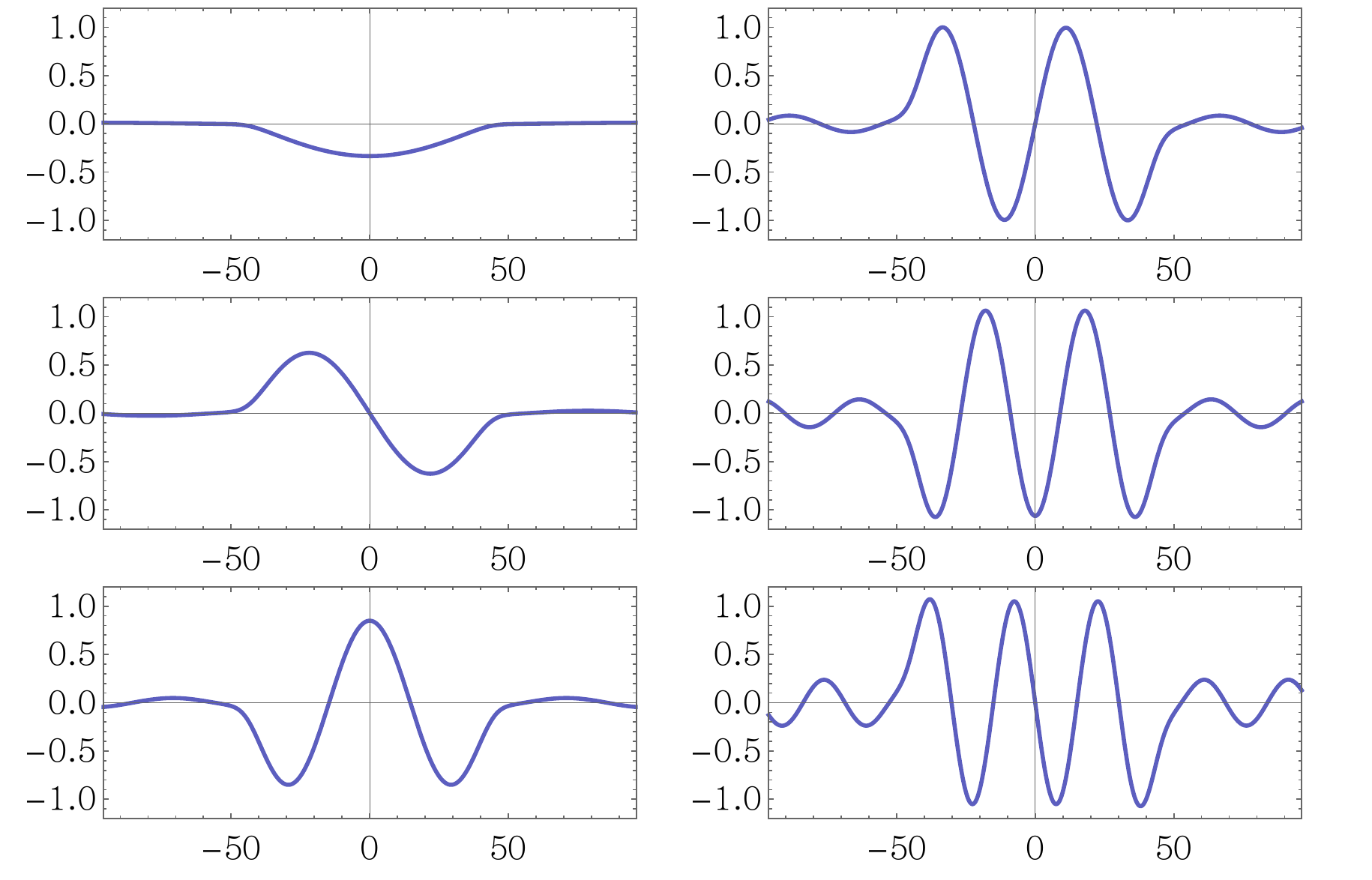}
	\caption{Quasinormal modes $\PPsi_{ln}(r_*)$ of the double P\"oschl-Teller potential as a function of $r_*/M$ with $n=1,2,3$ (left) and $n=4,5,6$ (right) for $l=1$ and $\lambda^2=10^{-10}$.}
	\label{QNM1}
\end{figure}

In Table \ref{QNFstatic}, we show the QNFs computed using both methods for the Damour-Solodukhin wormhole with $\lambda=10^{-5}$ and $l=0,1$. A graphic representation of the exact frequencies is provided in Fig.\ \ref{QNFplot1}. In  Fig.\ \ref{QNM1}, we plot the quasinormal modes $\PPsi_{ln}(r_*)$ for $l=1$ and various $n$ computed from the double  P\"oschl-Teller potential.
\begin{table*}[t]
		\centering
		\begin{tabular}{|c|c|c|c|c|}
			\hline
			Mode $n$&$M\omega_n (l=0$, P\"oschl-Teller )& $M\omega_n (l=0$, numerical)& $M\omega_n (l=1$, P\"oschl-Teller )&$M\omega_n (l=1$, numerical)\\
			\hline
			1& $0.03230 -1.666 \cdot 10^{-4}i$& $0.03210 -1.083 \cdot 10^{-4}i$&$0.03616-7.280\cdot 10^{-7}i$&$0.03501-1.962\cdot 10^{-8}i$\\
			2&$0.06427-7.413\cdot 10^{-4}i$ &$0.06382-5.720\cdot 10^{-4}i$&$0.07196-3.831\cdot 10^{-6}i$&$0.06982-4.390\cdot 10^{-7}i$\\
			3 &$0.09576-1.919\cdot 10^{-3}i$ & $0.09503-1.679\cdot 10^{-3}i$&$0.1071-1.309\cdot 10^{-5}i$&$0.1043-3.304\cdot 10^{-6}i$\\
			4& $0.1268-3.910\cdot 10^{-3}i$& $0.1258-3.709\cdot 10^{-3}i$&$0.1415-3.911\cdot 10^{-5}$&$0.1382-1.618\cdot 10^{-5}i$\\
			5& $0.1578-6.787\cdot 10^{-3}i$& $0.1566-6.739\cdot 10^{-3}i$&$0.1751-1.095\cdot 10^{-4}i$&$0.1715-6.266\cdot 10^{-5}i$\\
			6& $0.1888-0.01042i$& $0.1877-0.01057i$&$0.2077-2.908\cdot 10^{-4}i$&$0.2041-2.072\cdot 10^{-4}i$\\
			7& $0.2201-0.01456i$& $0.2192-0.01490i$&$0.2394-7.256\cdot 10^{-4}i$&$0.2358-5.999\cdot 10^{-4}i$\\
			8& $0.2517-0.01898i$& $0.2510-0.01948i$&$0.2701-1.662\cdot 10^{-3}i$&$0.2666-1.515\cdot 10^{-3}i$\\
			9& $0.2835-0.02355i$& $0.2832-0.02418i$&$0.3000-3.405\cdot 10^{-3}i$&$0.2968-3.295\cdot 10^{-3}i$\\
			10& $0.3155-0.02818i$& $0.3158-0.02900i$&$0.3295-6.145\cdot 10^{-3}i$&$0.3267-6.167\cdot 10^{-3}i$\\
			\hline
		\end{tabular}
		\caption{Scalar QNFs of the Damour-Solodukhin wormhole with $l=0,1$ and $\lambda^2=10^{-10}$. The result of the numerical computation for the exact potential is compared to the one obtained by replacing this with a P\"oschl-Teller potential.}
		\label{QNFstatic}
	\end{table*}

The P\"oschl-Teller approach gives a good approximation to the real part of the frequencies. The imaginary part is also in reasonable agreement, except for the smaller $n$'s. For $\lambda^2=10^{-10}$, we get $L=93.6 M$ and the real part of the frequencies is roughly proportional to $\pi/L\approx 0.0335/M$. More precisely, as expected from \req{ttto}, the QNFs are proportional to $\pi/L(1+ {\rm Im} \,R_{\rm BH}'(0)/L)$, whose value depends slightly on $l$: $0.9572\times \pi/L\approx 0.03211/M$ for $l=0$ and $1.0427\times \pi/L\approx 0.03498/M$ for $l=1$. This effect is smaller as we increase $L$, i.e., as we decrease $\lambda$, so the real part of the frequencies is almost independent of $l$. On the contrary, the imaginary part changes by several order of magnitudes for different $l$. Hence, for a fixed $n$, perturbations with large $l$ are much longer lived than those with low $l$.

\subsection{Time dependence and echoes}
We turn now to the time domain governed by eq.\ \req{timeeq}. We solve this equation numerically for an initial Gaussian perturbation near one of the photon spheres of the wormhole. We expect the primary signal received by an external observer for a perturbation of this kind to be very similar, and indeed almost identical, to the signal that would be produced by a black hole. The infalling wave in turn gives rise to a series of echoes consisting of a combination of wormhole QNMs. 

\subsubsection*{Numerical waveform and spectral analysis}
\begin{figure}[ht!]
\centering 
	\includegraphics[width=8.9cm]{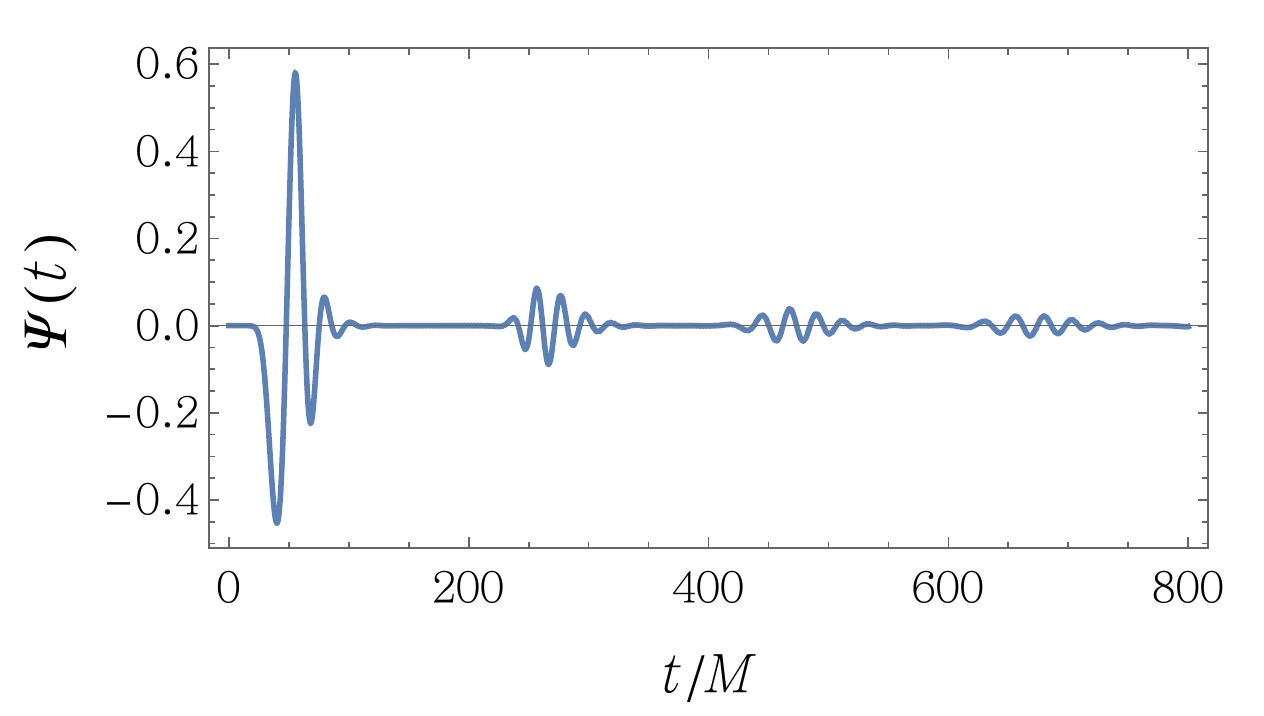}
	\caption{Time signal obtained from numerical integration of \req{timeeq}. We take $x$ to be sufficiently far away from the throat. The initial condition is a Gaussian perturbation near the photon sphere ($r\sim 3 M$) of the wormhole.}
	\label{TimeSignal}
\end{figure}

\begin{figure}[ht!]
\centering 
	\includegraphics[width=8.9cm]{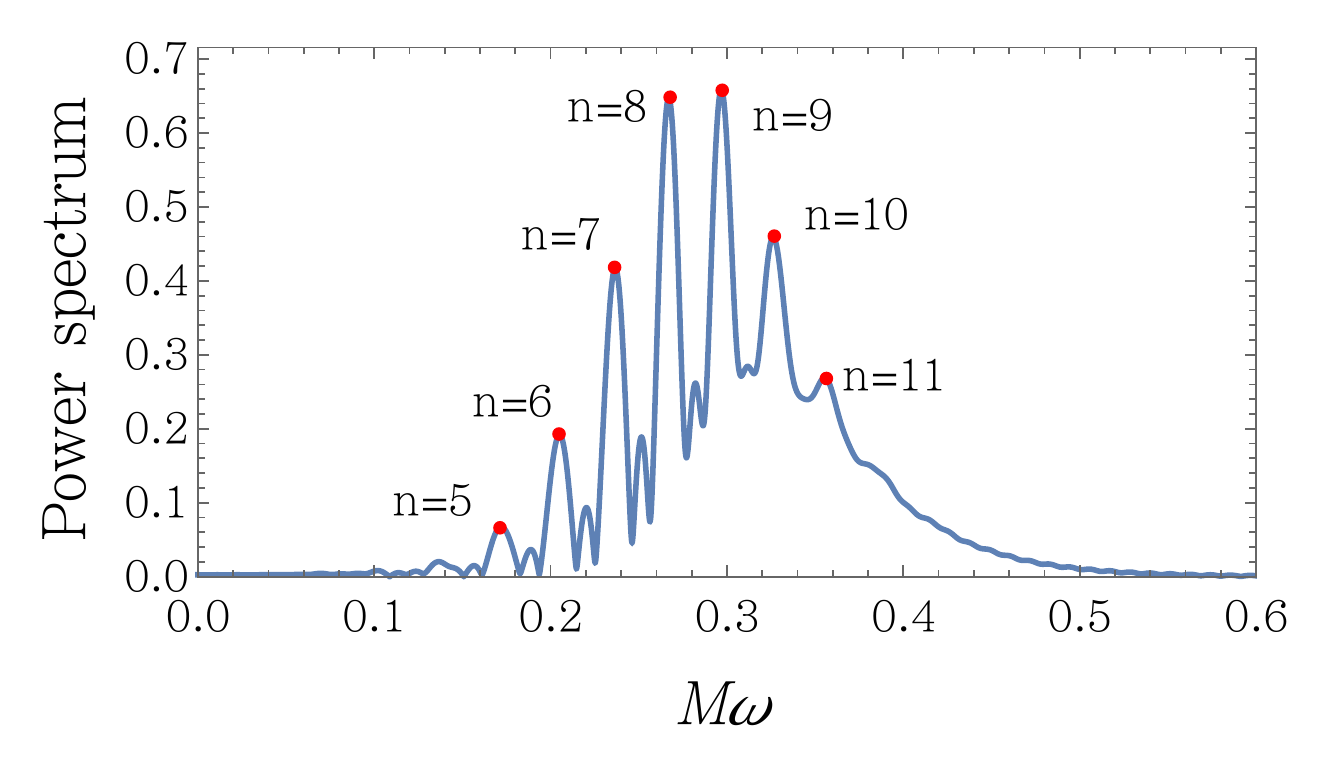}
	\caption{Power spectrum of the echoes waveform for $l=1$ (primary signal subtracted). The peaks, labelled by $n$, correspond to the QNFs of the wormhole, $\omega_n$. The absence of the $n=1,2,3,4$ modes is related to the fact that in this example the wormhole QNF which is closest to the black hole QNF dominating the primary signal corresponds to $n=9$. Due to the Gaussian character of the perturbation, only QNFs close to $\omega_9$ appear in the spectrum. }
	\label{Power}
\end{figure}

Fig.\ \ref{TimeSignal} shows the time signal obtained after numerical integration of eq.\ \req{timeeq}. As anticipated we first detect a signal which is essentially the one of a black hole, followed by a series of echoes roughly separated by a distance $2L\sim 190 M$.
Since it is the same signal that a black hole would produce, the primary wave is composed of the black hole QNMs. The leading frequency for $l=1$ is $(0.2929-0.09766 i)M^{-1}$, which appears as the main contribution to the primary signal. 
A Fourier analysis of the echo-part of the waveform determines the real part of the frequencies featuring in the echoes. The resulting power spectrum is shown in Fig.\ \ref{Power}. The most prominent peaks occur at the following frequencies: $M\omega=\{0.1721,\, 0.2051,\,0.2368,\,0.2671,\,0.2969,\,0.3263,\,0.3555\}$. Table \ref{QNFstatic} shows that these numbers are, with high accuracy, the QNFs of the wormhole. The labels in the plot show which QNF corresponds to each peak.
Hence, after the primary signal, the waveform is composed of wormhole QNMs.

As anticipated the waveforms of the individual echoes resemble each other. Moreover, the dominant frequency in the echo signal is $M\omega_9=0.2968-3.295\cdot 10^{-3}i$ which is the closest one to the black hole QNF dominating the primary signal, i.e. ${\rm Re}\, \omega^{\rm BH}=0.2929/M$. Thus we see that the echo waveform consists of a superposition of wormhole QNMs whose frequencies are concentrated around the dominant frequency of the black hole primary signal. This also explains the damping of the echoes: after the first echo, the amplitude of successive echoes should decrease roughly by a factor of $e^{{\rm Im}\,\omega_{n{\rm max}}2L}$, which in this case is $e^{{\rm Im}\,\omega_{9}2L}\approx 0.54$\footnote{This estimation is expected to work better for subleading echoes. By looking at the maxima of $|\Psi(t)|$, we can compare the amplitude of the primary signal with the one of the first echo, the latter with the one of the second, and so on. We find that this drops by factors: $\sim 0.15$, $\sim 0.43$ and $\sim 0.59$ respectively. }. Finally, the QNF spectrum explains the observation of Cardoso and Pani \cite{Cardoso:2017cqb,Cardoso:2017njb} that the frequency content of the echoes is decreasing with time: the QNMs with larger real frequency have larger imaginary frequency, so they decay faster than those of low frequency. The late time signal will be therefore composed of low frequencies corresponding, approximately, to the first few multiples of $\pi/L$.

\subsubsection*{Waveform reconstruction}


We can write the echo signals as in eq.\ \req{TSs}, with the coefficients $c_n$ given by \eqref{coeff2}, and thereby verify the accuracy of our method to reconstruct the echoes waveform. 

\begin{figure}[ht!]
	\centering 
	\includegraphics[width=8.9cm]{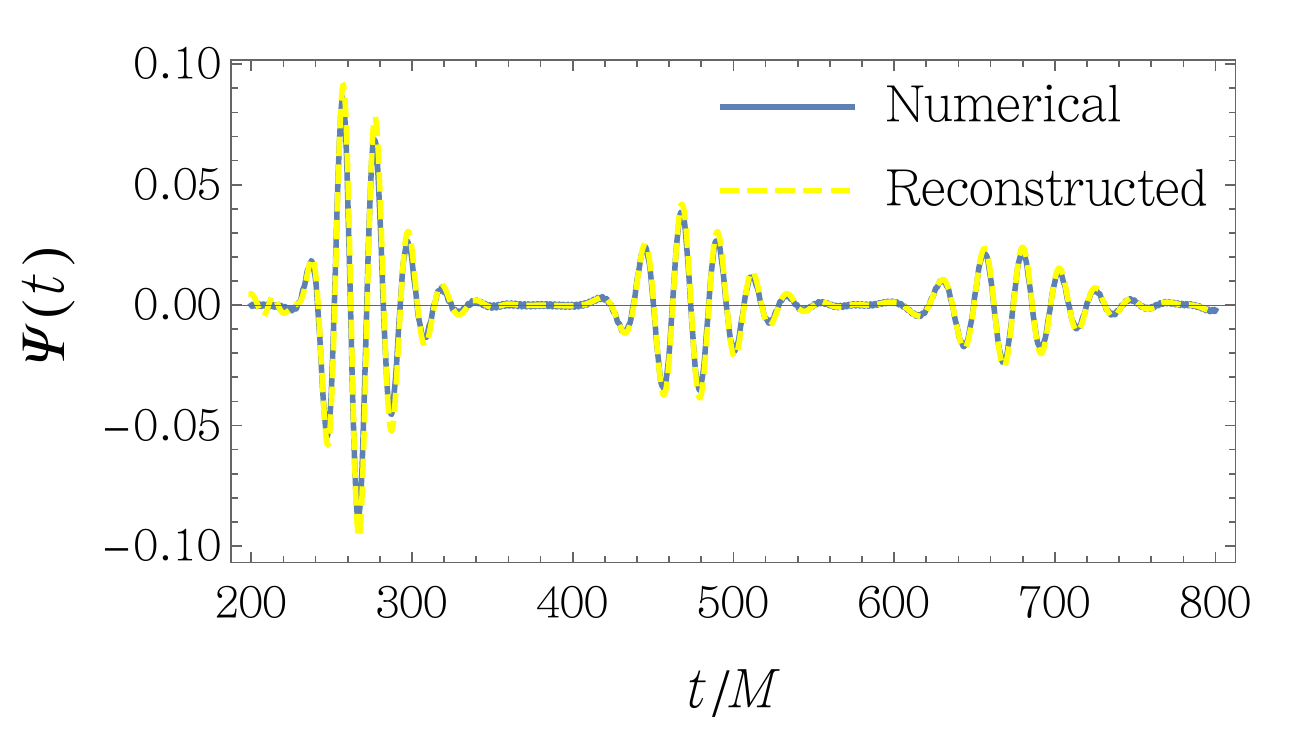}
	\caption{Echoes waveform: numerical computation (blue) and reconstructed (dashed)  asymptotic time signal. The second is obtained using eq. \req{TSs}. }
	\label{Recons}
\end{figure}
In Fig.\ \ref{Recons} we show the wave generated this way and compare it to the numerical one. They are in excellent agreement. This method allows us therefore to reconstruct the full echo waveform once the first echo and the QNFs spectrum are identified. 

\begin{figure}[ht!]
	\centering 
	\includegraphics[width=8.9cm]{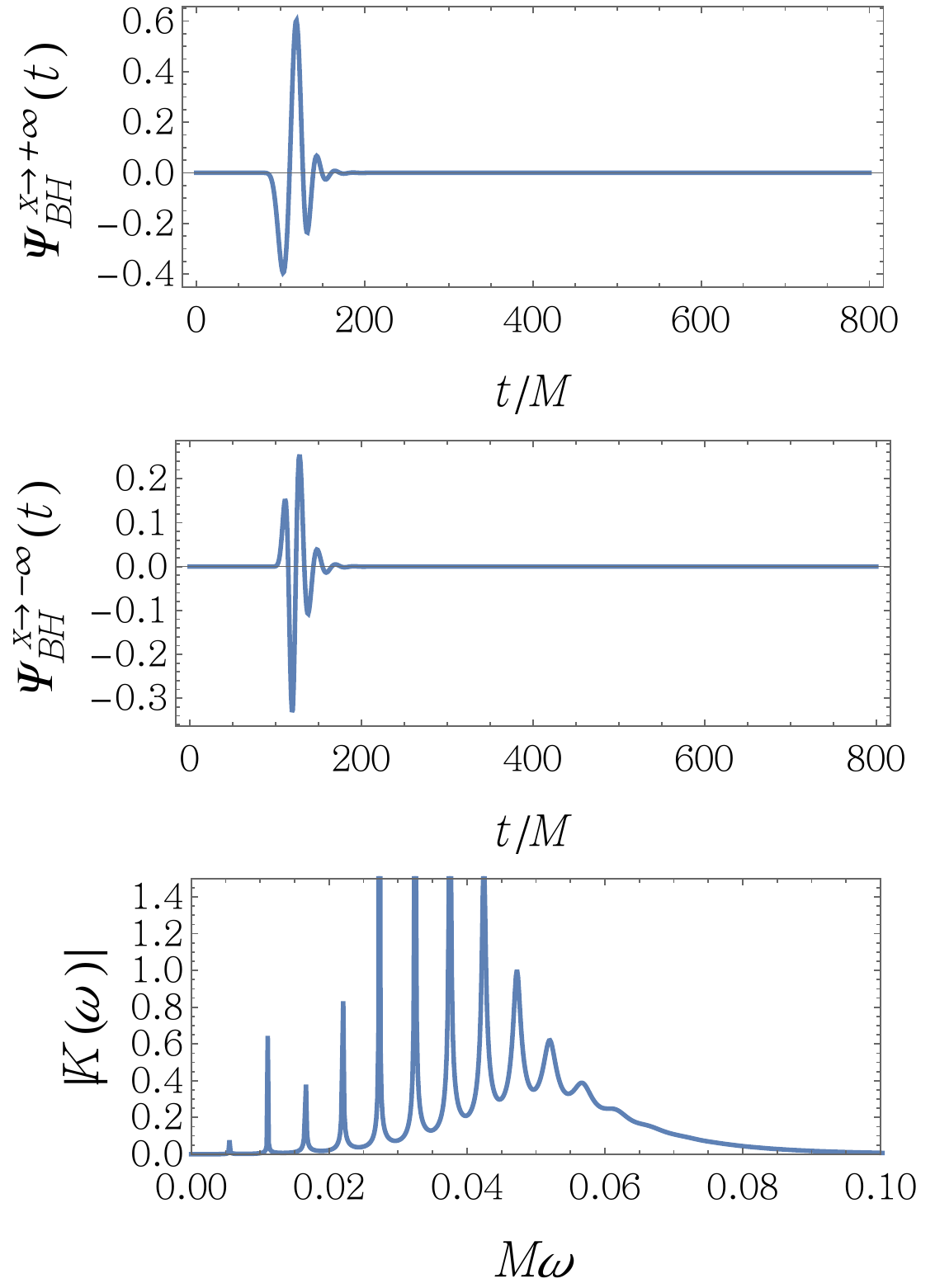}
	\caption{Top: asymptotic waveform corresponding to the single-bump Schwarzschild potential. Middle: near-horizon waveform for such potential. Bottom: transfer function computed using the transmission and reflection coefficients of the single-bump potential, as defined in Eq.\ \req{trans}.}
	\label{Recons2}
\end{figure}

\begin{figure}[ht!]
	\centering 
	\includegraphics[width=8.9cm]{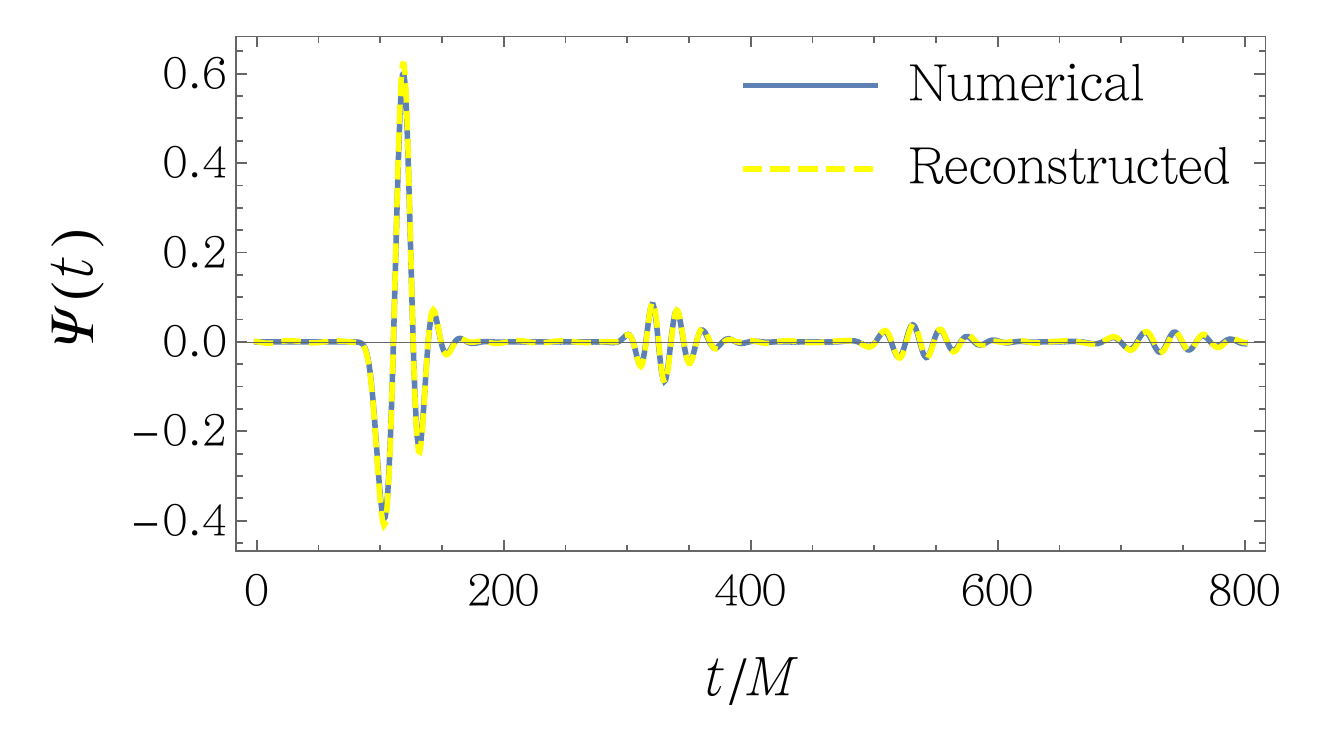}
	\caption{Reconstructed waveform (dashed) obtained using Mark et al's method \cite{Mark:2017dnq} according to Eq.\ \req{marktime} with $\omega_0=0$ versus exact numerical signal (blue).}
	\label{Recons3}
\end{figure}

Let us now consider the method of Mark et al \cite{Mark:2017dnq}. As we explained in Section \ref{markos}, this allows for a reconstruction of the full asymptotic waveform using eq.\ \req{marktime} from the knowledge of the near-horizon and asymptotic waveforms corresponding to the corresponding black hole potential along with its transfer function \req{trans}. In Fig.\ \ref{Recons2}, we plot the black hole waveforms for the same Gaussian perturbation considered before, as well as the absolute value of the transfer function, which displays singularities at the wormhole QNFs. Then, in Fig.\ \ref{Recons3}, we use eq.\  \req{marktime} to reconstruct the full wormhole asymptotic signal. The agreement is again excellent.

\section{Kerr-like wormhole}\label{KW}

Whatever might be the objects producing the gravitational waves attributed to merging black holes, they will have a non-zero spin. Hence we now consider phenomenologically more interesting Kerr-like wormhole as a toy model for a rotating ECO. 

We can construct a wormhole starting with the Kerr metric by performing a modification similar to what Damour and Solodukhin did for a Schwarzschild black hole \cite{Damour:2007ap}.  In particular, we consider the metric
\begin{align}\notag
&ds^2=-\left(1-\frac{2M r}{\Sigma}\right)dt^2-\frac{4Mar\sin^2\theta}{\Sigma}dtd\phi+\frac{\Sigma}{\hat \Delta}dr^2\\ \label{keeerr}
&+\Sigma d\theta^2+\left(r^2+a^2+\frac{2Ma^2r\sin^2\theta}{\Sigma}\right)\sin^2\theta d\phi^2\, ,
\end{align}
where
\begin{equation}
\Sigma\equiv r^2+a^2\cos^2\theta,\quad \hat\Delta \equiv r^2-2M(1+\lambda^2)r+a^2.
\end{equation}
Here, $M$ and $aM$ are the mass and the angular momentum of the would-be black hole respectively. If we set $\lambda^2=0$, the usual Kerr metric is recovered, but for any non-vanishing $\lambda^2$, the spacetime structure is completely different. The largest root of $\hat\Delta$ gives us the position of a special surface: $r_{+}=(1+\lambda^2)M+\sqrt{M^2(1+\lambda^2)^2-a^2}$. This is not a horizon anymore; indeed, the points with $r<r_{+}$ do not even exist. Instead, the surface $r=r_{+}$ is the throat of a wormhole which connects two asymptotically flat regions. This can be easily verified by performing the change of variable $r=r_{+}+\rho^2/M$. When written in terms of $\rho$, the previous metric is regular everywhere, and by allowing $\rho$ to take arbitrary real values, we see that the spacetime consists of two copies glued at $r=r_+$.
Since this happens for any $\lambda\neq 0$, by choosing a sufficiently small value we get a wormhole which is quasi-indistinguishable from a Kerr black hole as seen from the outside. The main difference would be the presence of echoes in the time signals produced by perturbations falling into the wormhole.

\subsection{Scalar perturbations} 
Let us consider a test scalar field $\Psi$ satisfying the Klein-Gordon equation $\Box \Psi=0$ in the background \req{keeerr}. Separating variables \cite{Teukolsky:1973ha},
\begin{equation}
\Psi(t,\vec{x})=\frac{1}{2\pi}\int d\omega e^{-i\omega t}\sum_{l=0}^{\infty}\sum_{m=-l}^{l}e^{im\phi}S_{lm}(\cos\theta)R_{lm}(r)\, ,
\end{equation}
 we find
\begin{align}
&\frac{d}{du}\left[(1-u^2)\frac{d}{du}\right]S_{lm}\\ \notag &+\left(A_{lm}+u^2a^2\omega^2-\frac{m^2}{1-u^2}\right)S_{lm}=0\, ,
\label{Spheroidal} \\
&\left(\frac{d^2}{dr_{*}^2}-V_{lm}(r,\omega)\right)\mathcal{R}_{lm}=-\omega^2\mathcal{R}_{lm}\, ,
\end{align}
where $u\equiv \cos\theta$, $A_{lm}= A_{lm}(a\omega)$ are the angular separation constants and $\mathcal{R}_{lm}\equiv \sqrt{r^2+a^2}R_{lm}$. The potential is
\begin{equation}
\begin{aligned}
V_{lm}&(r,\omega)=(r^2+a^2)^{-2}\Bigg[\Delta(A_{lm}+a^2\omega^2)+4rMa\omega m\\
&-a^2m^2+\frac{a^2-2r^2}{(r^2+a^2)^2}\Delta\hat \Delta+\frac{r}{2(r^2+a^2)}(\Delta\hat\Delta)'\Bigg]\, ,
\end{aligned}
\end{equation}
where: $'\equiv \partial/\partial r$,
$\Delta\equiv r^2-2Mr+a^2$, and the tortoise coordinate $r_*$ is defined through
\begin{equation}
\frac{dr_{*}}{dr}=\frac{r^2+a^2}{\sqrt{\Delta\hat\Delta}}\, .
\end{equation}
Note that the only difference with respect to the Kerr black hole is the appearance of $\hat \Delta$ instead of $\Delta$. This is particularly important in the definition of the tortoise coordinate. As happened for the Schwarzschild-like wormhole, for every value of $r$ there are two values of $r_{*}$. Using the auxiliary coordinate $\rho$ such that $r=r_{+}+\rho^2/M$, we can write 
\begin{equation}
r_{*}(\rho)=2M\int_{0}^{\rho}d\rho'\frac{(r_{+}+\rho'^2/M)^2+a^2}{\sqrt{(\rho'^2+\delta_1)(\rho'^2+\delta_2)(\rho'^2+\delta_3)}}\, ,
\label{tortoiserot}
\end{equation}
where
\begin{align}\notag
\delta_1&=\lambda^2M^2+M\sqrt{M^2(1+\lambda^2)^2-a^2}-M\sqrt{M^2-a^2},\\ \notag
\delta_2&=\lambda^2M^2+M\sqrt{M^2(1+\lambda^2)^2-a^2}+M\sqrt{M^2-a^2},\\
\delta_3&=2M\sqrt{M^2(1+\lambda^2)^2-a^2}\, .
\end{align}
Since $\delta_{1,2,3}>0$ (at least for $|a|<(1+\lambda^2) M$), there is a one-to-one relation between $r_{*}$ and $\rho$.

Note that the equation for the spin-weighted spheroidal harmonics $S_{lm}(u)$ does not depend on $\lambda^2$. So we can use the results available in the literature for Kerr black holes to deal with eq.\ \req{Spheroidal}. In particular, in order to solve the radial equation we will need to compute the scalar angular separation constants $A_{lm}(a\omega)$. We will be considering frequencies satisfying $a\omega\ll1$, so we can consider the series expansion of $A_{lm}$ in powers of $(a\omega)$:
\begin{equation}
A_{lm}(a\omega)=\sum_{n=0}^{\infty}f_n (a\omega)^n\, .
\end{equation}
The coefficients of this expansion can be found in \cite{Berti:2005gp}. In the remainder or the section, we use this series expansion up to $n=4$, which is more than enough to provide great accuracy in all our computations.

If $\lambda\ll1$, the potential $V_{lm}(r)$ becomes nearly identical to the one of a Kerr black hole, $V_{lm}(r)\approx V_{lm}^{\rm BH}(r) $, the most relevant difference being that $r(r_*)$ is a two-to-one function. Therefore, as a function of $r_*$, the potential develops two bumps. 
By \req{tortoiserot}, this potential is symmetric with respect to $r_*=0$. Now, for a Kerr black hole, the tortoise coordinate is
\begin{equation}
r_*^{\rm BH}=r+M \log \left(\frac{\Delta }{4 M^2}\right)-\frac{ M^2}{\sqrt{M^2-a^2}}\log \left(\frac{r-r_-}{r-r_+}\right)\, ,
\end{equation}
so that the potential for the black hole is $V_{lm}^{\rm BH}(r(r_*^{\rm BH}))$.
Just like for the Damour-Solodukhin wormhole, in the limit $\lambda\rightarrow 0$ and for fixed $r$, both coordinates $r_*$ and $r_*^{\rm BH}$ have the same behavior, but they differ by a constant (taking the positive branch of $r_*$):
\begin{equation}
r_*\rightarrow r_*^{\rm BH}+\frac{L}{2}\, .
\end{equation}
When $\lambda\ll1$, this distance is large and the leading term is
\begin{equation}
L=4M Z\log\left(\frac{2}{\lambda^2 Z}\right)+\mathcal{O}(1)\, ,
\end{equation}
where
\begin{equation}
Z=\frac{1}{2}\left(1+\frac{M}{\sqrt{M^2-a^2}}\right)\, .
\end{equation}
The separation between bumps depends on the wormhole angular momentum $a M$. When $a\rightarrow M$, $L$ diverges. For large separation $L$, the wormhole potential is approximated by
\begin{equation}
V_{lm}(r_*)=\theta(r_*)V_{lm}^{\rm BH}(r_*-L/2)+(r_*\leftrightarrow -r_*)\, ,
\end{equation}
where $V_l^{\rm BH}(\cdot)$ is the black hole potential as a function of the black hole tortoise coordinate  $r_*^{\rm BH}$.
Naturally, $V_{lm}(r_*)$ takes the form considered in Section \ref{DP}. 

\subsection{Quasinormal frequencies}\label{QNFKL}
\begin{figure}[t!]
	\centering 
	\includegraphics[width=8.5cm]{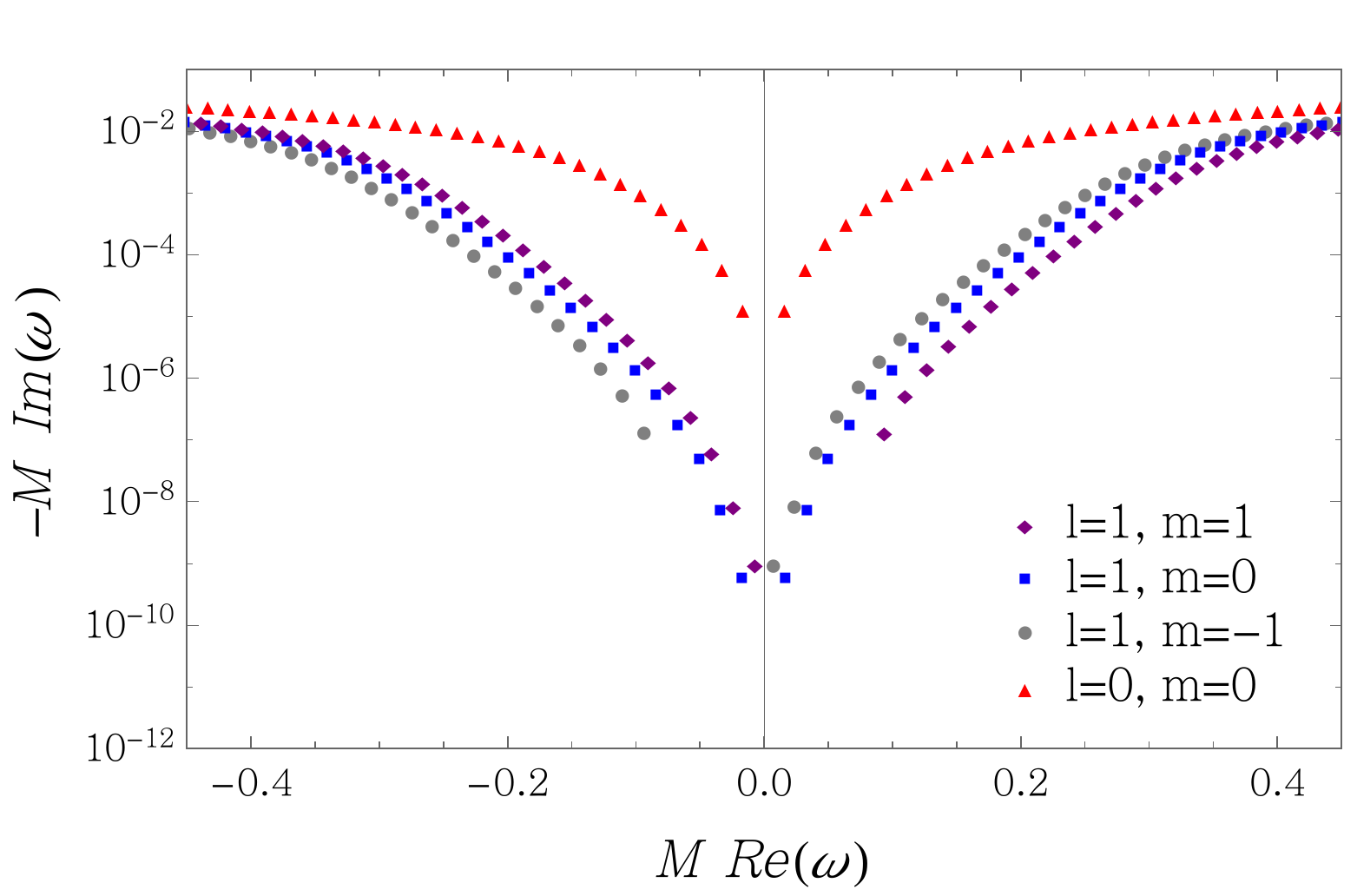}
	\includegraphics[width=8.5cm]{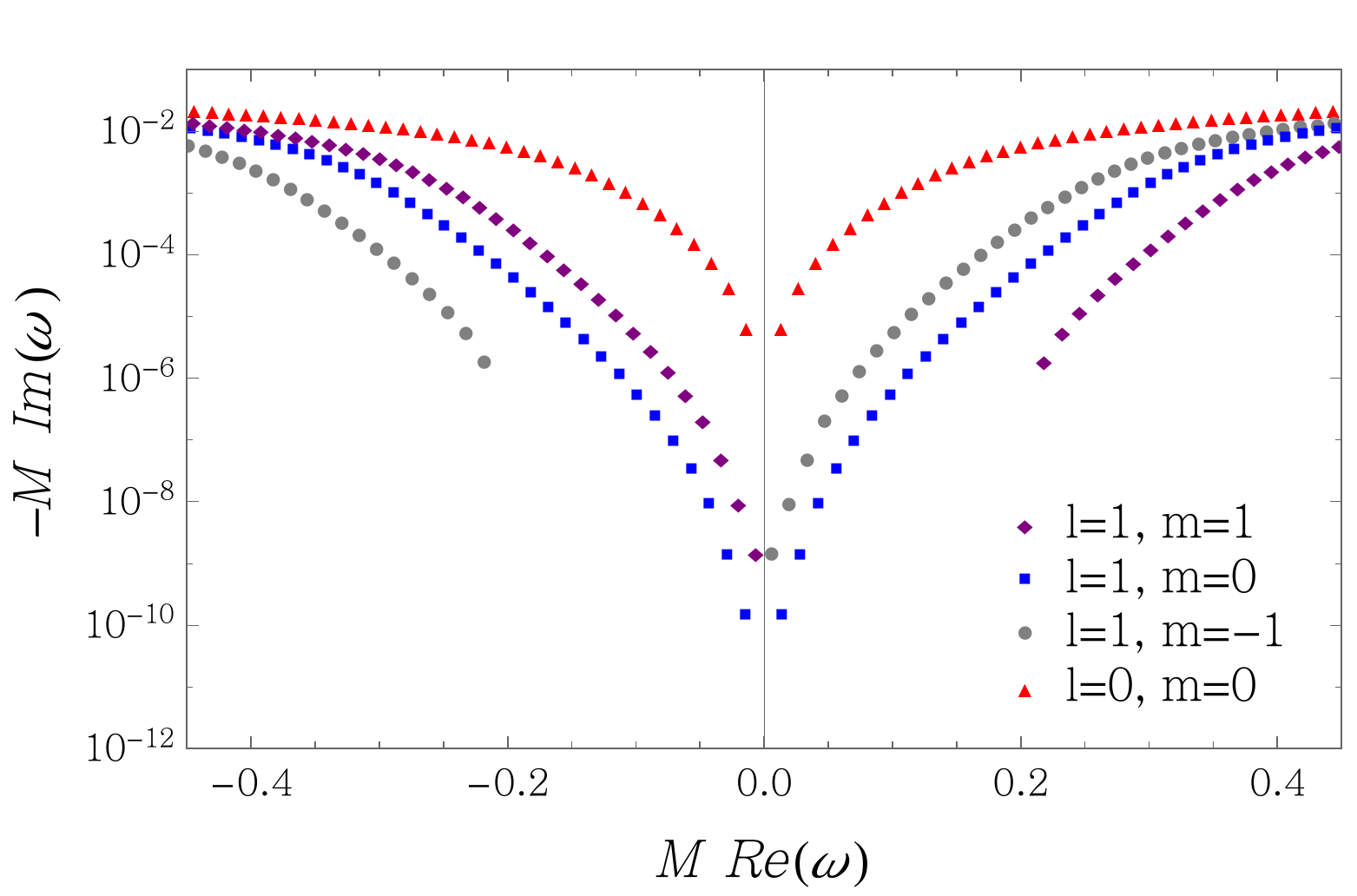}
	\includegraphics[width=8.5cm]{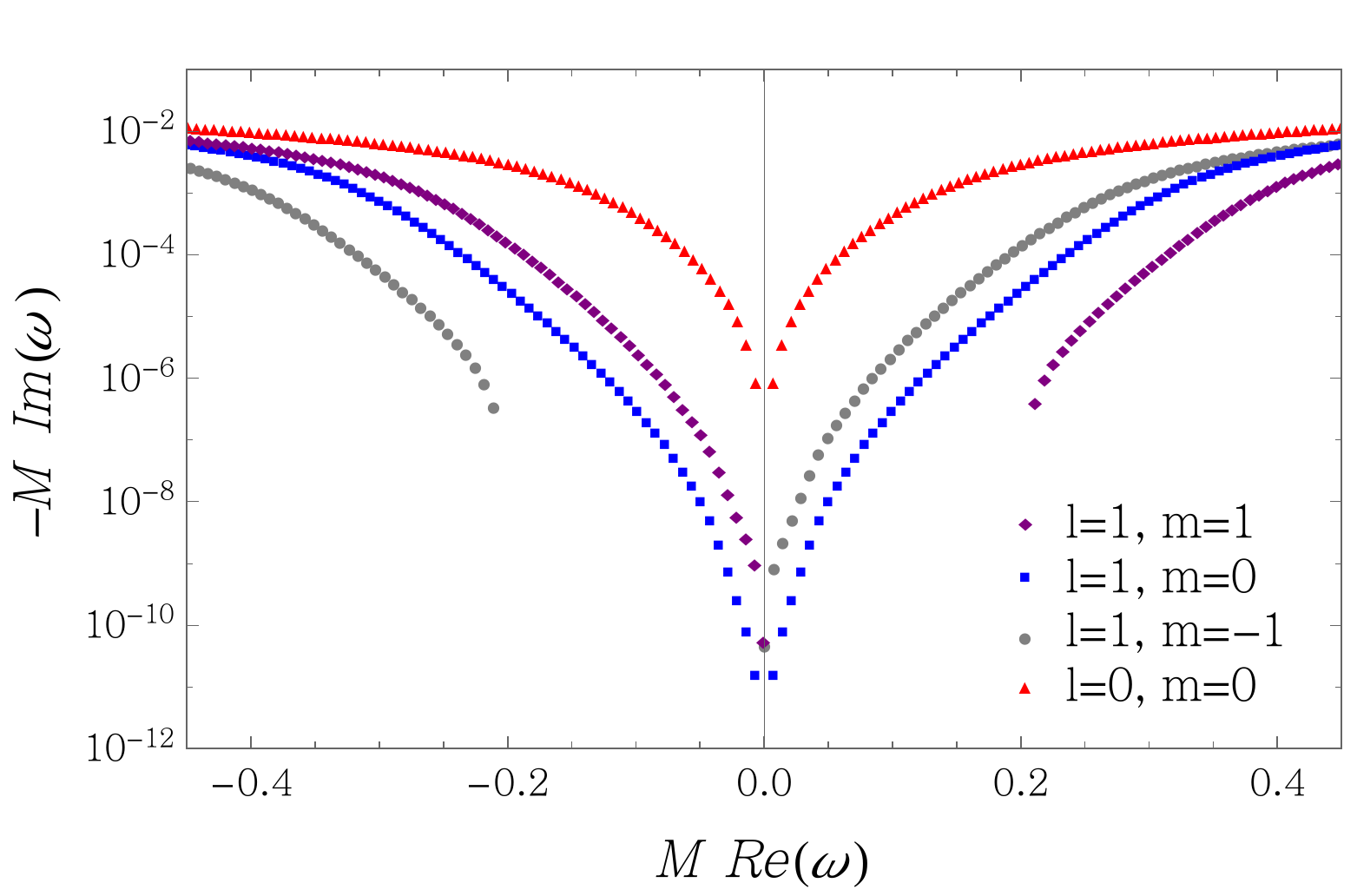}
	\caption{Scalar QNFs for the Kerr-wormhole. Top: Scalar QNFs with $l=0$, and $l=1$, $m=-1,0,1$ in a rotating wormhole with $\lambda^2=10^{-20}$ and $a=0.3M$. Middle: The same QNFs  for $\lambda^2=10^{-20}$ and $a=0.7M$. We can observe that the degeneration in $m$ is broken and that the difference increases with $a$. Bottom: QNFs for $\lambda^2=10^{-40}$ and $a=0.7M$. As $\lambda$ decreases, the spacing between QNFs decreases.  }
	\label{QNFm}
\end{figure}
%
%
In order to find the QNFs of the rotating wormhole, we need to determine the reflection coefficient of the black hole potential $V_{lm}^{\rm BH}(r_*^{\rm BH})$, and then solve eq.\ \req{refle}. A new phenomenon appears with respect to the static wormhole considered in the previous section. While the potential $V_{lm}^{\rm BH}(r_*^{\rm BH})$ still goes to zero at spatial infinity, it will no longer do so near the horizon.  As $r_*^{\rm BH}\rightarrow -\infty$ we get $\omega^2-V_{lm}^{\rm BH}(r_*^{\rm BH})\rightarrow (\omega-\omega_0)^2$, where
\begin{equation}\label{omeg}
\omega_0=\frac{a m}{2M r_+}\, .
\end{equation}
In other words, the black hole potential tends to a frequency-dependent plateau $U_0=\omega_0(2\omega-\omega_0)$, where $\omega_0$ is given in eq.\ \req{omeg} --- see Fig.\ \ref{fig1}. 
Therefore, the reflection coefficient of the black hole potential, $R_{\rm BH}$, should be determined by solving the following problem:
\begin{align}\label{Int}
&\mathcal{R}_{lm}=\\ \notag &\begin{cases}
T_{\rm BH}(\omega) e^{i\omega r_*^{\rm BH}}  \quad  &{\rm for}\quad r_*^{\rm BH}\rightarrow+ \infty\, ,\\
R_{\rm BH}(\omega) e^{-i(\omega-\omega_0) r_*^{\rm BH}}+e^{i(\omega-\omega_0) r_*^{\rm BH}}\quad  &{\rm for}\quad r_*^{\rm BH}\rightarrow-\infty\, .
\end{cases}
\end{align}
For a given $\omega$, the reflection coefficient $R_{\rm BH}(\omega)$ can be found numerically from \req{Int}. Then, eq.\ \req{refle} can be solved using, for instance, the secant method. In order to proceed, let us consider two wormholes with $a=0.7M$, and $\lambda^2=10^{-20}$ and $\lambda^2=10^{-40}$, respectively. Note that the distribution of possible spin parameters for stellar-mass black holes resulting from the merger of two has been observed to universally peak, precisely, at $a\sim0.7M$ \cite{Fishbach:2017dwv}, and hence our choice. Also, note that any $\lambda^2\neq 0$ would be related to some new-physics scale $l_0$ through $\lambda^2\sim l_0/M$. If $M\sim 10 M_{\odot}$, then $\lambda^2=10^{-40}$ corresponds roughly to $l_0\sim 0.1 \ell_{\rm P}$, while for $\lambda^2=10^{-20}$, this is approximately $10^{-16}$ meters, or about 1 GeV$^{-1}$.  The throats of these two wormholes have lengths $L=451 M$ and $L=230 M$, respectively.

Eqs. \req{ttto} and \req{tttt}, respectively, yield estimates of the real and imaginary parts of the leading QNFs. In particular, the approximation ${\rm Re} \, \omega_n\approx \omega_0+\frac{n\pi}{L}$ turns out to work quite accurately not only for $L\gg M$ (corresponding to $\lambda^2\ll1$), but also for  $L/(n\pi)\sim M$. In particular, for $\lambda^2=10^{-40}$, $a=0.7M$, $l=m=1$, we find numerically: ${\rm Re} \, \omega_{\{0,1,2\}}=\{0.20418,0.2113,0.21833 \}$, while the approximation yields: $\omega_0+\{0,1,2\}\pi/L=\{0.20418,0.2115,0.21812 \}$, which are very close to the exact values. But also, for example, for $n=50$, we find ${\rm Re} \, \omega_{50}=0.5513$, whereas  $\omega_0+50 \pi/L=0.5525$, which only differs by $\sim 0.2\%$. With respect to the imaginary parts of the frequencies closest to $\omega_0$, numerical evidence suggest that for $m=0$ they scale as $1/L^3$, while for $m\neq 0$ they go with $1/L^2$, when $L$ is large enough. This would be in agreement with the expansion \req{ttt}, if we take into account that ${\rm Re}\left[R'_{\rm BH}(\omega_0)\right]\neq 0$ for $m\neq 0$. A full understanding of this dependence requires further investigation though.

 In Fig.\ \ref{QNFm}, we show the QNFs of the wormholes mentioned above for $l=0$ and $l=1$. We also show the QNFs for a wormhole with $a=0.3M$ and $\lambda^2=10^{-20}$ in order to observe how these get modified as the angular momentum changes. We note immediately that a Zeeman-like splitting of the QNFs appears, characteristic of the presence of rotation \cite{Berti:2009kk}, which breaks the degeneracy for different values of $m$. 
 
In addition, there are some QNFs that do not appear in the plot because they have a positive imaginary part. These are the ones with ${\rm Re}(\omega)({\rm Re}(\omega)-\omega_0)<0$.  These modes are unstable, as they grow exponentially with time, and appear when $|R_{\rm BH}(\omega)|>1$. This instability is reminiscent of black hole superradiance \cite{Brito:2015oca} and is argued to occur for any horizonless rotating compact object with an ergosphere --- see \eg \cite{Friedman:1978wla,Cardoso:2007az,Maggio:2017ivp}. The expected endpoint of such an `ergoregion instability' would be a slowly-spinning compact object, surrounded by a spinning particle cloud. Applied to our rotating wormhole, we see that there are no unstable modes for $|\omega_0|<\pi/L$, which imposes the condition on the spin
\begin{equation}
\left|\frac{a}{M}\right|<\frac{4\frac{\pi M}{|m|L}}{1+4\left(\frac{\pi M}{|m|L}\right)^2}\, .
\end{equation} 
This formula means that if the spin is below this quantity, modes with $|m'|\le |m|$ are all stable. Obviously, if we want modes of arbitrary $m$ to be stable, $a$ must vanish. However, as $a$ decreases, fewer modes are unstable. One expects that the angular momentum will effectively cease to decrease before reaching $a=0$. The resulting object would then be slowly-spinning, probably with $a \sim M^2/L$. Note however that this instability does not prevent the compact object --- possibly produced after the merger of two objects --- from having a large angular momentum initially.

\subsection{Signal reconstruction}

In section \ref{DSW}, we observed that the echo waveform for a static Schwarzschild-like wormhole was completely formed by a combination of its QNMs, the most relevant being the one whose QNF was closest to the main black hole QNF, dominating the primary signal. We expect the same behavior to occur for rotating wormholes. After a perturbation, say, near the photon sphere is produced, we expect to detect a primary signal corresponding to the one a rotating black hole would produce. This will contain the QNMs of the black hole. Later on, echoes composed of wormhole QNMs will appear. Since the primary signal contains the black hole QNFs, the echoes will try to imitate this frequency content --- something they can do because the wormhole QNMs nearly form a Fourier basis, as we explained. 
 
In Section \ref{DSW} we showed that the method put forward in Section \ref{ECHOE} for reconstructing all the subleading echoes from the first one worked very accurately for the static Damour-Solodukhin wormhole. Let us now apply this method to our Kerr-like wormhole. We do not solve the time-domain equation directly but rather we model the leading-echo waveform $\Psi^{(0)}_{\rm {1^{\rm st}}echo}(t)$ as a Gaussian wave packet of the form given in eq.\ \req{WP}, for some frequency $\omega^{\rm BH}_{lm0}$ that corresponds to the leading black hole QNF controlling the primary signal.

\begin{figure}[ht!]
	\centering 
	\includegraphics[width=8.6cm]{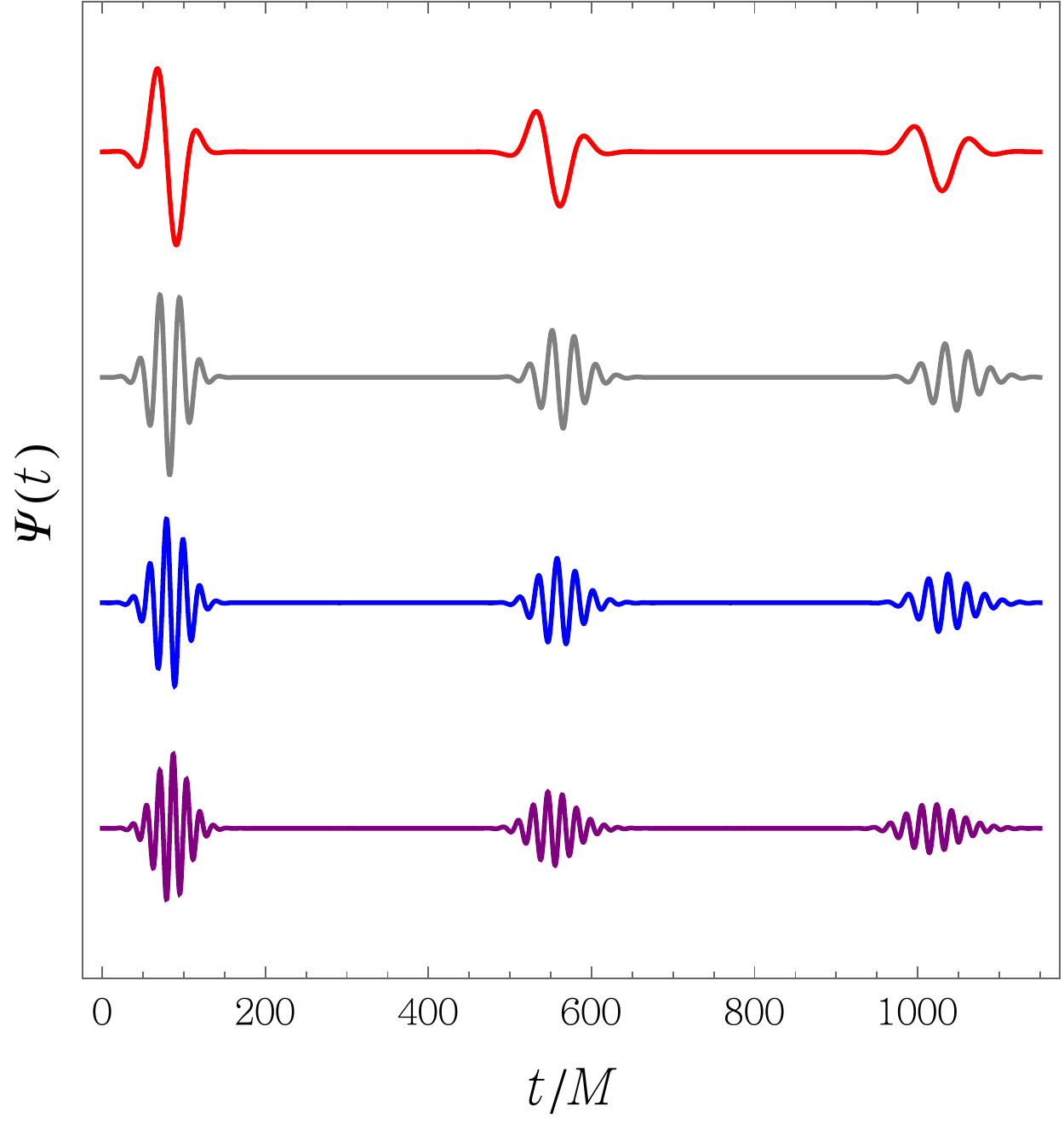}
	\caption{Construction of echo waveforms from Gaussian wave packets in a rotating wormhole with $a=0.7 M$ and $\lambda^2=10^{-20}$. From top to bottom: $(l,m)=(0,0),\, (1,-1),\, (1,0),\, (1,1).$ In all panels, the first echo fits a Gaussian with the corresponding leading QNF of a Kerr black hole with $a=0.7 M$ --- see Eq.\ \req{WP}. From this initial wave packet we obtain the coefficients of the wormhole QNM expansion from Eq.\ \req{ccc}, and we construct the rest of the signal using Eq.\ \req{TSS}. The echoes appear with a period $T\sim 2L\approx 460 M$, and the damping is approximately independent of $l$ and $m$.}
	\label{waveforms}
\end{figure}

More explicitly, the full echo waveform that an observer would detect can be written as
\begin{equation}
\Psi(t)=\sum_{l=0}^{\infty}\sum_{m=-l}^{l}\sum_{n=-\infty}^{\infty}c_{lmn}e^{-i\omega_{lmn} t}\, ,
\label{TSS}
\end{equation}
where the coefficients are given as
\begin{equation}\label{ccc}
c_{lmn}=\frac{1}{2L}\int_0^{2L}dt \Psi^{(0)}_{\rm {1^{\rm st}}echo} e^{i\omega_{lmn} t}\, .
\end{equation}
Modeling the first echo as
\begin{equation}
\Psi^{(0)}_{lm}(t)=e^{-i\omega^{\rm BH}_{lm0}(t-t_0)}e^{-\frac{(t-t_0)^2}{2\tau^2}}\, ,
\label{WP2}
\end{equation}
where $\omega^{\rm BH}_{lm0}$ is the leading QNF of the black hole for the given $l$ and $m$, and substituting the actual QNFs of the wormhole and the ones of the would-be black hole, we get a waveform which should resemble the actual waveform produced by a Gaussian perturbation. For example, for a Kerr black hole with $a=0.7$ the leading scalar quasinormal frequencies for $l=0,1$ are \cite{Berti:2009kk}: $M\omega^{\rm BH}_{000}=0.1140-0.09863 i\, ,M\omega^{\rm BH}_{1-10}=0.2519-0.0955 i\, ,M\omega^{\rm BH}_{100}=0.3031-0.09243 i\, ,M\omega^{\rm BH}_{110}=0.3792-0.08885 i$ . We can also use the wormhole QNFs for $\lambda^2=10^{-20}$ computed previously. The resulting waveforms for $\tau=20M$ are shown in Fig.\ \ref{waveforms}.

For all values $(l,m)$ shown, the first echo fits perfectly the Gaussian wave packet \req{WP}, but it is constructed with the quasinormal frequencies of the wormhole. The echoes appear after a period of $2L\approx 460 M$ because the real parts of the frequencies are spaced approximately $\pi/L$. Damping and deformation occur as a consequence of the QNFs being complex and not exactly equispaced. Interestingly, the damping seems to be mostly independent of $l$ and $m$. Fig.\ \ref{QNFm} shows that modes with larger $l$ and with larger $am$ decay slower. The reason why we see similar decay is due to the fact that also the main frequency of the wave packet --- the corresponding QNF of the black hole --- changes, having a larger real part for larger $l$ and larger $am$. Hence, both effects approximately cancel out, rendering the damping of the echoes roughly equal for every $m$ and $l$.

Another important observation is that we do not see any unstable modes in the echo waveform, at least at early times. This can be related to the fact that the leading black hole QNF is sufficiently far from the frequencies of the unstable modes, so that these are barely excited and consequently they get a tiny amplitude. In addition, these unstable modes have very small (positive) imaginary parts, so they take a long time to grow. For example, for  $l=1$, $m=1$, we have $M\omega^{\rm BH}_{110}=0.3792-0.08885 i$, while the first unstable QNF (corresponding to $n=-1$) of the wormhole with $\lambda^2=10^{-20}$ is  $M\omega_{11-1}=0.1901 + 7.929\cdot10^{-7} i$. The timescale at which the instability becomes observable is of the order $t_{\rm inst}\sim \frac{\log(|c_n|^{-1})}{{\rm Im}(\omega_n)}$, which is very large since ${\rm Im}(\omega_n)$ and $c_n$ are tiny. In particular, for the Gaussian wave packet \req{WP2}, we can use the coefficients \req{WPC} to get
\begin{equation}
t_{\rm inst}\sim\frac{\tau^2{\rm Re}(\omega^{\rm BH}_{lm0}-\omega_{lmn})^2}{2{\rm Im}(\omega_{lmn})}\, .
\end{equation}
For the example at hand, this gives $t_{\rm inst}\sim 10^7 M$. 

The take-home message is that we can construct a realistic echo waveform for the wormhole by knowing its QNFs and the black hole ones. The first echo, together with the knowledge of the QNFs, can be used to reconstruct the rest of the signal. Naturally, the QNFs can be in principle determined once the reflection coefficient of the single-bump potential is known, using eq.\ \req{refle}. The modeling of the first echo from a given primary signal using \req{WP2} will not be, however, perfect in general, and a more refined method would be desirable. This can be achieved using Mark et al's construction. In that construction, one requires the knowledge of the near-horizon, would-be black hole waveform, as well as the transfer function, in addition to the asymptotic primary  signal.

From the point of view of an asymptotic observer, our method and that of Mark et al are comparably powerful and useful. Gravitational-wave detectors cannot collect all the QNFs of the compact object, nor can they assess directly what the near-horizon response of a black hole would be with the same mass as a given ECO. The asymptotic primary waveform of an astrophysical compact  object, resulting after a merger for example, can be split in two parts: the prompt ringdown response and the echo-part of the signal. The prompt ringdown can be studied separately, and combined with data from the merger process that formed the compact object, would give a very good estimate of the mass and spin of the end product. This can be used to determine the black hole with the same mass and charge, such that all black hole data are known (QNFs, reflection coefficient\ldots). However, this is not enough information to predict the detailed waveform of the ensuing echoes, as the prompt response is to a very good approximation insensitive to the kind of ECO. One first needs to measure at least the first echo, to either reconstruct the rest of the signal following our method, or to reconstruct the transfer function $\cal K$ used in the approach of Mark et al. This means that modeled searches in future gravitational-wave data analysis are most powerful when the details of the first echo are captured by one or more free parameters related to the type of ECO.

\section{Conclusions}\label{conclusions}

We have studied the quasinormal spectrum of perturbations in static and stationary wormhole spacetimes, how such a spectrum connects to asymptotic time-signals produced, and how the echoes in such signals can be reconstructed from the primary signal governed by the corresponding effective black hole potential or from the leading echo. We have applied our results to Damour and Solodukhin's Schwarzschild-like wormhole \cite{Damour:2007ap}, and to a new Kerr-like wormhole.  A detailed summary of our results can be found in the Introduction.

With regards to future work, it would be interesting to clarify whether the general properties of the wormhole QNFs observed here --- such as the approximate spacing by $\pi/L$ of the real parts, or the fact that the wormhole QNF that dominates echoes is the one which is closest to the black hole QNF controlling the primary signal --- extend to more general ECOs. 
It would also be desirable to obtain a more quantitative understanding of the consequences of the instability observed for the rotating wormholes. Our study has been restricted to probe scalar fields on wormhole backgrounds. An analysis of gravitational perturbations would be interesting, but this requires embedding the corresponding wormhole metrics in an explicit theory. Examples of wormhole  solutions in certain modified theories of gravity are available --- see e.g., \cite{Harko:2013yb,Lobo:2009ip}, although these generally differ significantly from Schwarzschild or Kerr away from the would-be horizon. Attempts at constructing pathology-free wormhole solutions to Einstein gravity coupled to non-exotic matter have also been carried out in \cite{Ayon-Beato:2015eca,Canfora:2017gno}, so those could in principle provide  candidates for such explorations.

Naturally, it would be interesting to further study perturbations in other alternatives to black holes such as  boson stars~\cite{Berti:2006qt,Macedo:2013qea,Macedo:2013jja}, gravastars~\cite{Chirenti:2016hzd}, other wormholes~\cite{Konoplya:2016hmd,Nandi:2016uzg} and quantum-corrected objects~\cite{Barcelo:2017lnx,Brustein:2017koc}. Many of these horizonless black hole mimickers and ECOs that are commonly studied share some of the symmetries of the black hole spacetime. This has led to candidate smoking-gun signals which include short lifetimes for highly spinning objects due to an ergoregion instability and, as we have seen, gravitational-wave echoes spaced at relatively short times. However some of the black hole alternatives that have emerged in string theory break several or all of the black hole's symmetries. An important but under-explored issue is to assess to what extent the analysis and the signals studied here apply to such solutions.

A particular example are fuzzballs \cite{Mathur:2005zp,Bena:2007kg,Skenderis:2008qn,Mathur:2008nj,Balasubramanian:2008da}, which are horizonless and typically highly asymmetric solutions. Fuzzballs may not share the smoking gun signals  expected from more symmetric solutions and discussed here. This is because fuzzballs do not have integrable and thus separable geodesic equations. Hence, radially infalling modes will mix with angular modes, possibly leading to an enhancement of the time needed for perturbations to leak out \cite{Bena:2017upb}, and hence the time between echoes. One also expects that the inner structure of fuzzballs strongly modifies the relation between the different echoes signals, and between the first echo and the primary signal. The formation process itself of fuzzballs involves a series of quantum transitions which go together with novel types of gravitational wave bursts \cite{Hertog:2017vod}. Evidently it would be very interesting to have a better understanding of their waveform. We will return to these open questions in future work \cite{Hertog_upcoming}.


\acknowledgments
We thank Vitor Cardoso and Kwinten Fransen for enlightening discussions. The work of PB was supported by a postdoctoral fellowship from the National Science Foundation of Belgium (FWO). The work of PAC was supported by a ``la Caixa Severo
Ochoa'' International pre-doctoral grant and in part by the Spanish Ministry of Science and
Education grants FPA2012-35043-C02-01 and FPA2015-66793-P, and by the Centro de Excelencia Severo
Ochoa Program grant SEV-2012-0249. PAC also thanks the ``Visiting Graduate Fellowship" program of Perimeter Institute.
This work was also supported in part by the European Research Council grant ERC-2013-CoG 616732 HoloQosmos, by the C16/16/005 grant of the KU Leuven, and by the FWO grant G092617N. We also acknowledge networking support by the COST Action GWverse CA16104.

\onecolumngrid  
\appendix

\section{Analytic QNFs for simple symmetric double potentials}\label{sdp}

In this appendix, we discuss analytic approximations to the QNFs of wormhole-like potentials. We consider three simple examples: two Dirac-delta peaks, a double square well and a combination of two P\"oschl-Teller potentials.

\subsubsection*{Double Dirac-delta potential}
The simplest possible example corresponds to a double Dirac-delta potential $V^{\delta\delta}$, for which $U(x)$ takes the form
\begin{equation}
U^{\delta}(x)= \nu \delta(x)\,,
\end{equation}
where the parameter $\nu$ measures the strength of the potential. If we wanted to use the double Dirac-delta to approximate the QNFs of some other potential $V_1(x)$, $\nu$ could be related to the defining parameters of such potential through
\begin{equation}\label{vvv}
2\nu=\int_{-\infty}^{+\infty} V_1(x)dx\, .
\end{equation}

The QNFs condition \req{refle} translates into
\begin{equation}
e^{-i \omega_n L}=\frac{e^{-i \pi n}}{\left(1-\frac{2i \omega_n}{\nu}\right)}\, ,
\end{equation}
whose general solution is
\begin{equation}\label{Lambert}
\omega_m=\frac{i}{ L}\left[W_{\{m\}}\left(e^{-i \pi n} \frac{\nu L}{2}e^{\frac{\nu  L}{2}}\right)-\frac{\nu L}{2}\right]\, , \,\, m \in \mathbb{Z}\, ,
\end{equation}
where $W_{\{m\}}$ is the $m$-th branch of the Lambert function. The leading frequency corresponds to: $m=-1$, $n=1$, the subleading one to: $m=-1$, $n=2$, the next to: $m=-2$, $n=3$, and so on.
Using \req{ttt}, it is easy to see that the leading QNFs can be quickly estimated from the simple expression
\begin{equation}\label{lds}
\omega_n= \frac{\pi n }{L}  \left[\left(1-\frac{2 }{\nu L}+\dots \right)-i\left( \frac{2\pi n}{\nu^2L^2}+\dots \right)\right]\, ,
\end{equation}
which yields very accurate results for the first QNFs when compared with \req{Lambert}. Observe that while the real parts are of order $\sim 1/L$, the imaginary ones are approximately proportional to $1/L^3$, which illustrates their relative smallness, as argued in general in section \ref{QNMs}. A more precise approximation for the imaginary parts is obtained using Eq.\ \req{tttt}.

\subsubsection*{Double rectangular barrier}\label{Doublerec}
Another simple potential consists of two rectangular barriers of height $V_0$ and width $a$, 
\begin{equation}
U^{\sqcap}(x)=V_0 \left[\theta(x)-\theta(x-a) \right] \, ,
\end{equation}
If we wanted to use $V^{\sqcap\sqcap}(x)$ to approximate another potential $V_1(x)$, the parameters $a$ and $V_0$ could be fixed by imposing
\begin{align}
V_0&=V_1(x^{\rm max})\, ,\\ \label{aa}
a&=\frac{1}{2V_1(x^{\rm max})}\int_{-\infty}^{+\infty}V_1(x) dx\, .
\end{align}
Alternatively, $a$ could be also related to some average width of the single bumps in $V_1(x)$, which should yield a result similar to the one given by \req{aa}.

A textbook calculation yields the reflection coefficient of the double-rectangular barrier. From this, using \req{refle}, we obtain the following expression for the QNFs of $V^{\sqcap\sqcap}(x)$
\begin{equation}\label{dfd}
e^{-i\omega_n L}= -\frac{e^{-i \pi n} V_0}{2i \omega_n \sqrt{\omega_n^2-V_0} \cot (a\sqrt{\omega_n^2-V_0})+(2\omega_n^2-V_0)}\, .
\end{equation}
Using \req{ttto} and \req{tttt}, we can again obtain quick estimations for the QNFs: 
\begin{equation}\label{ldm}
\omega_n= \frac{\pi n }{L}  \left[\left(1-\frac{2 }{\sqrt{V_0} L }\coth(a\sqrt{V_0})+\dots \right)-i\left( \frac{2\pi n}{V_0 L^2}\csch^2(a \sqrt{V_0}) +\dots \right)\right]\, ,
\end{equation}
which again yields very accurate results for the first QNFs when compared with the ones obtained solving \req{dfd} numerically. Note again that the imaginary parts are $\mathcal{O}(L^{-3})$. Eq.\ \req{ldm} reduces to the double Dirac-delta result \req{lds} in the limit $V_0/a\gg 1$ if we identify $a V_0\equiv \nu$.

\subsubsection*{Double P\"oschl-Teller potential}\label{DoublePT}
Let us finally consider a double P\"oschl-Teller potential $V^{\rm PT}$, derived from the standard form \cite{Poschl:1933zz},
\begin{equation}
U^{\rm PT}(x)=V_0\cdot {\rm sech}^2(\alpha x)\, .
\end{equation}
The P\"oschl-Teller potential $U^{\rm PT}(x)$ provides a more sophisticated approximation to black hole potentials than the previous two,  while still allowing for analytic calculations. --- it is commonly used to calculate approximate QNFs of black hole spacetimes.  If we wanted to use the resulting double-bump potential $V^{\rm PT}$ to mimic a given potential $V_1(x)$, we could do that by relating $V_0$ to the height of $V_1$ at its maximum, and $\alpha$ to its second derivative as
\begin{align}
V_0&=V_1(x^{\rm max})\, ,\\ \notag
\alpha^2&=-\frac{1}{2V_1(x^{\rm max})}\left. \frac{d^2V_1}{dx^2}\right|_{x^{\rm max}}\,.
\end{align}

The QNMs solving \req{QNMproblem} for the P\"oschl-Teller  potential $U^{\rm PT}(x)$ are
\begin{equation}
\begin{aligned}
&\PPsi_U(x)=A(\tanh (\alpha x)+1)^{\frac{i \omega_n}{2 \alpha}} (1-\tanh (\alpha x))^{-\frac{i \omega_n}{2\alpha}}   \cdot\, _2F_1\left[\xi ,1-\xi ; \frac{i\omega_n}{\alpha}+1;\frac{1}{2} (\tanh (\alpha x)+1)\right]\\
&+B\, 2^{\frac{i\omega_n}{\alpha}} (\tanh (\alpha x)+1)^{-\frac{i \omega_n}{2\alpha}} (1-\tanh (\alpha x))^{-\frac{i \omega_n}{2\alpha}}  \cdot\, _2F_1\left[\xi - \frac{i\omega_n}{\alpha},-\frac{i\omega_n}{\alpha}-\xi +1;1-\frac{i\omega_n}{\alpha};\frac{1}{2} (\tanh (\alpha x)+1)\right]\, ,
\end{aligned}
\end{equation}
where we defined
$\xi\equiv (1\pm i\sqrt{4V_0/\alpha^2-1})/2$. We use this result to approximate the QNMs of the Damour-Solodukhin wormhole using Eq.\ \req{eq:psi_from_psi_U} --- see Fig.\ \ref{QNM1}.

For a wave incoming from $-\infty$,  $A$ and $B$ must be related by $B=R^{\rm PT}(\omega) A$.
The expression for the QNFs of a double P\"oschl-Teller can be shown to be given by
\begin{equation}
e^{-i\omega_n L}=e^{-i \pi n} \frac{\Gamma(1+i\omega_n/\alpha)\Gamma(\xi-i\omega_n/\alpha)\Gamma(1-\xi-i\omega_n/\alpha)}{\Gamma(1-i\omega_n/\alpha)\Gamma(\xi)\Gamma(1-\xi)} \, .
\end{equation}
Now, using the following property of the digamma function:
\begin{equation}\label{zz}
{\rm Im } \left [\psi_0(z)+\psi_0(\bar{z})\right]=0 \quad \forall z \in \mathbb{C} \, ,
\end{equation}
it is possible to obtain the following approximations for the real and imaginary parts of the QNFs
\begin{equation}\label{ldm2}
\omega_n= \frac{\pi n }{L}  \left[\left(1+\frac{2\gamma_{\rm E}+\psi_0(1-\xi)+\psi_0(\xi) }{\alpha L}+\dots \right)-i\left( \frac{n\pi (\psi_0(1-\xi)+\psi_0(\xi)) ((\psi_0(1-\xi)+\psi_0(\xi))-1)}{2\alpha^2 L^2}+\dots\right)\right]\, ,
\end{equation}
where $\gamma_{\rm E}$ is the Euler-Mascheroni constant, and note that, as a consequence of Eq.\ \req{zz},
\begin{equation}
\psi_0(1-\xi)+\psi_0(\xi) \in \mathbb{R}\quad  \forall \,V_0,\alpha\, .
\end{equation}
Once, again, we observe that the imaginary part is $\mathcal{O}(L^{-3})$. The above formula reduces to the double Dirac-delta result \req{lds} in the limit $V_0/\alpha^2 \rightarrow 0$ if we identify $V_0/\alpha\equiv \nu/2$.

\renewcommand{\leftmark}{\MakeUppercase{Bibliography}}
\phantomsection
\bibliography{Gravities}

 \newcommand{\noop}[1]{}
\begin{thebibliography}{52}%
\makeatletter
\providecommand \@ifxundefined [1]{%
 \@ifx{#1\undefined}
}%
\providecommand \@ifnum [1]{%
 \ifnum #1\expandafter \@firstoftwo
 \else \expandafter \@secondoftwo
 \fi
}%
\providecommand \@ifx [1]{%
 \ifx #1\expandafter \@firstoftwo
 \else \expandafter \@secondoftwo
 \fi
}%
\providecommand \natexlab [1]{#1}%
\providecommand \enquote  [1]{``#1''}%
\providecommand \bibnamefont  [1]{#1}%
\providecommand \bibfnamefont [1]{#1}%
\providecommand \citenamefont [1]{#1}%
\providecommand \href@noop [0]{\@secondoftwo}%
\providecommand \href [0]{\begingroup \@sanitize@url \@href}%
\providecommand \@href[1]{\@@startlink{#1}\@@href}%
\providecommand \@@href[1]{\endgroup#1\@@endlink}%
\providecommand \@sanitize@url [0]{\catcode `\\12\catcode `\$12\catcode
  `\&12\catcode `\#12\catcode `\^12\catcode `\_12\catcode `\%12\relax}%
\providecommand \@@startlink[1]{}%
\providecommand \@@endlink[0]{}%
\providecommand \url  [0]{\begingroup\@sanitize@url \@url }%
\providecommand \@url [1]{\endgroup\@href {#1}{\urlprefix }}%
\providecommand \urlprefix  [0]{URL }%
\providecommand \Eprint [0]{\href }%
\providecommand \doibase [0]{http://dx.doi.org/}%
\providecommand \selectlanguage [0]{\@gobble}%
\providecommand \bibinfo  [0]{\@secondoftwo}%
\providecommand \bibfield  [0]{\@secondoftwo}%
\providecommand \translation [1]{[#1]}%
\providecommand \BibitemOpen [0]{}%
\providecommand \bibitemStop [0]{}%
\providecommand \bibitemNoStop [0]{.\EOS\space}%
\providecommand \EOS [0]{\spacefactor3000\relax}%
\providecommand \BibitemShut  [1]{\csname bibitem#1\endcsname}%
\let\auto@bib@innerbib\@empty
\bibitem [{\citenamefont {Abbott}\ \emph {et~al.}(2016)\citenamefont {Abbott}
  \emph {et~al.}}]{Abbott:2016blz}%
  \BibitemOpen
  \bibfield  {author} {\bibinfo {author} {\bibfnamefont {B.~P.}\ \bibnamefont
  {Abbott}} \emph {et~al.} (\bibinfo {collaboration} {Virgo, LIGO
  Scientific}),\ }\href {\doibase 10.1103/PhysRevLett.116.061102} {\bibfield
  {journal} {\bibinfo  {journal} {Phys. Rev. Lett.}\ }\textbf {\bibinfo
  {volume} {116}},\ \bibinfo {pages} {061102} (\bibinfo {year} {2016})},\
  \Eprint {http://arxiv.org/abs/1602.03837} {arXiv:1602.03837 [gr-qc]}
  \BibitemShut {NoStop}%
\bibitem [{\citenamefont {Lunin}\ and\ \citenamefont
  {Mathur}(2002)}]{Lunin:2001jy}%
  \BibitemOpen
  \bibfield  {author} {\bibinfo {author} {\bibfnamefont {O.}~\bibnamefont
  {Lunin}}\ and\ \bibinfo {author} {\bibfnamefont {S.~D.}\ \bibnamefont
  {Mathur}},\ }\href {\doibase 10.1016/S0550-3213(01)00620-4} {\bibfield
  {journal} {\bibinfo  {journal} {Nucl. Phys.}\ }\textbf {\bibinfo {volume}
  {B623}},\ \bibinfo {pages} {342} (\bibinfo {year} {2002})},\ \Eprint
  {http://arxiv.org/abs/hep-th/0109154} {arXiv:hep-th/0109154 [hep-th]}
  \BibitemShut {NoStop}%
\bibitem [{\citenamefont {Almheiri}\ \emph {et~al.}(2013)\citenamefont
  {Almheiri}, \citenamefont {Marolf}, \citenamefont {Polchinski},\ and\
  \citenamefont {Sully}}]{Almheiri:2012rt}%
  \BibitemOpen
  \bibfield  {author} {\bibinfo {author} {\bibfnamefont {A.}~\bibnamefont
  {Almheiri}}, \bibinfo {author} {\bibfnamefont {D.}~\bibnamefont {Marolf}},
  \bibinfo {author} {\bibfnamefont {J.}~\bibnamefont {Polchinski}}, \ and\
  \bibinfo {author} {\bibfnamefont {J.}~\bibnamefont {Sully}},\ }\href
  {\doibase 10.1007/JHEP02(2013)062} {\bibfield  {journal} {\bibinfo  {journal}
  {JHEP}\ }\textbf {\bibinfo {volume} {02}},\ \bibinfo {pages} {062} (\bibinfo
  {year} {2013})},\ \Eprint {http://arxiv.org/abs/1207.3123} {arXiv:1207.3123
  [hep-th]} \BibitemShut {NoStop}%
\bibitem [{\citenamefont {Marolf}(2017)}]{Marolf:2017jkr}%
  \BibitemOpen
  \bibfield  {author} {\bibinfo {author} {\bibfnamefont {D.}~\bibnamefont
  {Marolf}},\ }\href {\doibase 10.1088/1361-6633/aa77cc} {\bibfield  {journal}
  {\bibinfo  {journal} {Rept. Prog. Phys.}\ }\textbf {\bibinfo {volume} {80}},\
  \bibinfo {pages} {092001} (\bibinfo {year} {2017})},\ \Eprint
  {http://arxiv.org/abs/1703.02143} {arXiv:1703.02143 [gr-qc]} \BibitemShut
  {NoStop}%
\bibitem [{\citenamefont {Mazur}\ and\ \citenamefont
  {Mottola}(2001)}]{Mazur:2001fv}%
  \BibitemOpen
  \bibfield  {author} {\bibinfo {author} {\bibfnamefont {P.~O.}\ \bibnamefont
  {Mazur}}\ and\ \bibinfo {author} {\bibfnamefont {E.}~\bibnamefont
  {Mottola}},\ }\href@noop {} {\  (\bibinfo {year} {2001})},\ \Eprint
  {http://arxiv.org/abs/gr-qc/0109035} {arXiv:gr-qc/0109035 [gr-qc]}
  \BibitemShut {NoStop}%
\bibitem [{\citenamefont {Schunck}\ and\ \citenamefont
  {Mielke}(2003)}]{Schunck:2003kk}%
  \BibitemOpen
  \bibfield  {author} {\bibinfo {author} {\bibfnamefont {F.~E.}\ \bibnamefont
  {Schunck}}\ and\ \bibinfo {author} {\bibfnamefont {E.~W.}\ \bibnamefont
  {Mielke}},\ }\href {\doibase 10.1088/0264-9381/20/20/201} {\bibfield
  {journal} {\bibinfo  {journal} {Class. Quant. Grav.}\ }\textbf {\bibinfo
  {volume} {20}},\ \bibinfo {pages} {R301} (\bibinfo {year} {2003})},\ \Eprint
  {http://arxiv.org/abs/0801.0307} {arXiv:0801.0307 [astro-ph]} \BibitemShut
  {NoStop}%
\bibitem [{\citenamefont {Morris}\ \emph {et~al.}(1988)\citenamefont {Morris},
  \citenamefont {Thorne},\ and\ \citenamefont {Yurtsever}}]{Morris:1988tu}%
  \BibitemOpen
  \bibfield  {author} {\bibinfo {author} {\bibfnamefont {M.~S.}\ \bibnamefont
  {Morris}}, \bibinfo {author} {\bibfnamefont {K.~S.}\ \bibnamefont {Thorne}},
  \ and\ \bibinfo {author} {\bibfnamefont {U.}~\bibnamefont {Yurtsever}},\
  }\href {\doibase 10.1103/PhysRevLett.61.1446} {\bibfield  {journal} {\bibinfo
   {journal} {Phys. Rev. Lett.}\ }\textbf {\bibinfo {volume} {61}},\ \bibinfo
  {pages} {1446} (\bibinfo {year} {1988})}\BibitemShut {NoStop}%
\bibitem [{\citenamefont {Damour}\ and\ \citenamefont
  {Solodukhin}(2007)}]{Damour:2007ap}%
  \BibitemOpen
  \bibfield  {author} {\bibinfo {author} {\bibfnamefont {T.}~\bibnamefont
  {Damour}}\ and\ \bibinfo {author} {\bibfnamefont {S.~N.}\ \bibnamefont
  {Solodukhin}},\ }\href {\doibase 10.1103/PhysRevD.76.024016} {\bibfield
  {journal} {\bibinfo  {journal} {Phys. Rev.}\ }\textbf {\bibinfo {volume}
  {D76}},\ \bibinfo {pages} {024016} (\bibinfo {year} {2007})},\ \Eprint
  {http://arxiv.org/abs/0704.2667} {arXiv:0704.2667 [gr-qc]} \BibitemShut
  {NoStop}%
\bibitem [{\citenamefont {Mathur}(2005)}]{Mathur:2005zp}%
  \BibitemOpen
  \bibfield  {author} {\bibinfo {author} {\bibfnamefont {S.~D.}\ \bibnamefont
  {Mathur}},\ }\bibfield  {booktitle} {\emph {\bibinfo {booktitle} {{The
  quantum structure of space-time and the geometric nature of fundamental
  interactions. Proceedings, 4th Meeting, RTN2004, Kolymbari, Crete, Greece,
  September 5-10, 2004}}},\ }\href {\doibase 10.1002/prop.200410203} {\bibfield
   {journal} {\bibinfo  {journal} {Fortsch. Phys.}\ }\textbf {\bibinfo {volume}
  {53}},\ \bibinfo {pages} {793} (\bibinfo {year} {2005})},\ \Eprint
  {http://arxiv.org/abs/hep-th/0502050} {arXiv:hep-th/0502050 [hep-th]}
  \BibitemShut {NoStop}%
\bibitem [{\citenamefont {Holdom}\ and\ \citenamefont
  {Ren}(2017)}]{Holdom:2016nek}%
  \BibitemOpen
  \bibfield  {author} {\bibinfo {author} {\bibfnamefont {B.}~\bibnamefont
  {Holdom}}\ and\ \bibinfo {author} {\bibfnamefont {J.}~\bibnamefont {Ren}},\
  }\href {\doibase 10.1103/PhysRevD.95.084034} {\bibfield  {journal} {\bibinfo
  {journal} {Phys. Rev.}\ }\textbf {\bibinfo {volume} {D95}},\ \bibinfo {pages}
  {084034} (\bibinfo {year} {2017})},\ \Eprint
  {http://arxiv.org/abs/1612.04889} {arXiv:1612.04889 [gr-qc]} \BibitemShut
  {NoStop}%
\bibitem [{\citenamefont {Cardoso}\ \emph
  {et~al.}(2016{\natexlab{a}})\citenamefont {Cardoso}, \citenamefont
  {Franzin},\ and\ \citenamefont {Pani}}]{Cardoso:2016rao}%
  \BibitemOpen
  \bibfield  {author} {\bibinfo {author} {\bibfnamefont {V.}~\bibnamefont
  {Cardoso}}, \bibinfo {author} {\bibfnamefont {E.}~\bibnamefont {Franzin}}, \
  and\ \bibinfo {author} {\bibfnamefont {P.}~\bibnamefont {Pani}},\ }\href
  {\doibase 10.1103/PhysRevLett.117.089902, 10.1103/PhysRevLett.116.171101}
  {\bibfield  {journal} {\bibinfo  {journal} {Phys. Rev. Lett.}\ }\textbf
  {\bibinfo {volume} {116}},\ \bibinfo {pages} {171101} (\bibinfo {year}
  {2016}{\natexlab{a}})},\ \bibinfo {note} {[Erratum: Phys. Rev.
  Lett.117,no.8,089902(2016)]},\ \Eprint {http://arxiv.org/abs/1602.07309}
  {arXiv:1602.07309 [gr-qc]} \BibitemShut {NoStop}%
\bibitem [{\citenamefont {Cardoso}\ \emph
  {et~al.}(2016{\natexlab{b}})\citenamefont {Cardoso}, \citenamefont {Hopper},
  \citenamefont {Macedo}, \citenamefont {Palenzuela},\ and\ \citenamefont
  {Pani}}]{Cardoso:2016oxy}%
  \BibitemOpen
  \bibfield  {author} {\bibinfo {author} {\bibfnamefont {V.}~\bibnamefont
  {Cardoso}}, \bibinfo {author} {\bibfnamefont {S.}~\bibnamefont {Hopper}},
  \bibinfo {author} {\bibfnamefont {C.~F.~B.}\ \bibnamefont {Macedo}}, \bibinfo
  {author} {\bibfnamefont {C.}~\bibnamefont {Palenzuela}}, \ and\ \bibinfo
  {author} {\bibfnamefont {P.}~\bibnamefont {Pani}},\ }\href {\doibase
  10.1103/PhysRevD.94.084031} {\bibfield  {journal} {\bibinfo  {journal} {Phys.
  Rev.}\ }\textbf {\bibinfo {volume} {D94}},\ \bibinfo {pages} {084031}
  (\bibinfo {year} {2016}{\natexlab{b}})},\ \Eprint
  {http://arxiv.org/abs/1608.08637} {arXiv:1608.08637 [gr-qc]} \BibitemShut
  {NoStop}%
\bibitem [{\citenamefont {Barcelo}\ \emph {et~al.}(2011)\citenamefont
  {Barcelo}, \citenamefont {Garay},\ and\ \citenamefont
  {Jannes}}]{Barcelo:2010vc}%
  \BibitemOpen
  \bibfield  {author} {\bibinfo {author} {\bibfnamefont {C.}~\bibnamefont
  {Barcelo}}, \bibinfo {author} {\bibfnamefont {L.~J.}\ \bibnamefont {Garay}},
  \ and\ \bibinfo {author} {\bibfnamefont {G.}~\bibnamefont {Jannes}},\ }\href
  {\doibase 10.1007/s10701-011-9577-9} {\bibfield  {journal} {\bibinfo
  {journal} {Found. Phys.}\ }\textbf {\bibinfo {volume} {41}},\ \bibinfo
  {pages} {1532} (\bibinfo {year} {2011})},\ \Eprint
  {http://arxiv.org/abs/1002.4651} {arXiv:1002.4651 [gr-qc]} \BibitemShut
  {NoStop}%
\bibitem [{\citenamefont {Barcelo}\ \emph {et~al.}(2015)\citenamefont
  {Barcelo}, \citenamefont {Carballo-Rubio}, \citenamefont {Garay},\ and\
  \citenamefont {Jannes}}]{Barcelo:2014cla}%
  \BibitemOpen
  \bibfield  {author} {\bibinfo {author} {\bibfnamefont {C.}~\bibnamefont
  {Barcelo}}, \bibinfo {author} {\bibfnamefont {R.}~\bibnamefont
  {Carballo-Rubio}}, \bibinfo {author} {\bibfnamefont {L.~J.}\ \bibnamefont
  {Garay}}, \ and\ \bibinfo {author} {\bibfnamefont {G.}~\bibnamefont
  {Jannes}},\ }\href {\doibase 10.1088/0264-9381/32/3/035012} {\bibfield
  {journal} {\bibinfo  {journal} {Class. Quant. Grav.}\ }\textbf {\bibinfo
  {volume} {32}},\ \bibinfo {pages} {035012} (\bibinfo {year} {2015})},\
  \Eprint {http://arxiv.org/abs/1409.1501} {arXiv:1409.1501 [gr-qc]}
  \BibitemShut {NoStop}%
\bibitem [{\citenamefont {Abedi}\ \emph {et~al.}(2016)\citenamefont {Abedi},
  \citenamefont {Dykaar},\ and\ \citenamefont {Afshordi}}]{Abedi:2016hgu}%
  \BibitemOpen
  \bibfield  {author} {\bibinfo {author} {\bibfnamefont {J.}~\bibnamefont
  {Abedi}}, \bibinfo {author} {\bibfnamefont {H.}~\bibnamefont {Dykaar}}, \
  and\ \bibinfo {author} {\bibfnamefont {N.}~\bibnamefont {Afshordi}},\
  }\href@noop {} {\  (\bibinfo {year} {2016})},\ \Eprint
  {http://arxiv.org/abs/1612.00266} {arXiv:1612.00266 [gr-qc]} \BibitemShut
  {NoStop}%
\bibitem [{\citenamefont {Ashton}\ \emph {et~al.}(2016)\citenamefont {Ashton},
  \citenamefont {Birnholtz}, \citenamefont {Cabero}, \citenamefont {Capano},
  \citenamefont {Dent}, \citenamefont {Krishnan}, \citenamefont {Meadors},
  \citenamefont {Nielsen}, \citenamefont {Nitz},\ and\ \citenamefont
  {Westerweck}}]{Ashton:2016xff}%
  \BibitemOpen
  \bibfield  {author} {\bibinfo {author} {\bibfnamefont {G.}~\bibnamefont
  {Ashton}}, \bibinfo {author} {\bibfnamefont {O.}~\bibnamefont {Birnholtz}},
  \bibinfo {author} {\bibfnamefont {M.}~\bibnamefont {Cabero}}, \bibinfo
  {author} {\bibfnamefont {C.}~\bibnamefont {Capano}}, \bibinfo {author}
  {\bibfnamefont {T.}~\bibnamefont {Dent}}, \bibinfo {author} {\bibfnamefont
  {B.}~\bibnamefont {Krishnan}}, \bibinfo {author} {\bibfnamefont {G.~D.}\
  \bibnamefont {Meadors}}, \bibinfo {author} {\bibfnamefont {A.~B.}\
  \bibnamefont {Nielsen}}, \bibinfo {author} {\bibfnamefont {A.}~\bibnamefont
  {Nitz}}, \ and\ \bibinfo {author} {\bibfnamefont {J.}~\bibnamefont
  {Westerweck}},\ }\href@noop {} {\  (\bibinfo {year} {2016})},\ \Eprint
  {http://arxiv.org/abs/1612.05625} {arXiv:1612.05625 [gr-qc]} \BibitemShut
  {NoStop}%
\bibitem [{\citenamefont {Abedi}\ \emph {et~al.}(2017)\citenamefont {Abedi},
  \citenamefont {Dykaar},\ and\ \citenamefont {Afshordi}}]{Abedi:2017isz}%
  \BibitemOpen
  \bibfield  {author} {\bibinfo {author} {\bibfnamefont {J.}~\bibnamefont
  {Abedi}}, \bibinfo {author} {\bibfnamefont {H.}~\bibnamefont {Dykaar}}, \
  and\ \bibinfo {author} {\bibfnamefont {N.}~\bibnamefont {Afshordi}},\
  }\href@noop {} {\  (\bibinfo {year} {2017})},\ \Eprint
  {http://arxiv.org/abs/1701.03485} {arXiv:1701.03485 [gr-qc]} \BibitemShut
  {NoStop}%
\bibitem [{\citenamefont {Price}\ and\ \citenamefont
  {Khanna}(2017)}]{Price:2017cjr}%
  \BibitemOpen
  \bibfield  {author} {\bibinfo {author} {\bibfnamefont {R.~H.}\ \bibnamefont
  {Price}}\ and\ \bibinfo {author} {\bibfnamefont {G.}~\bibnamefont {Khanna}},\
  }\href@noop {} {\  (\bibinfo {year} {2017})},\ \Eprint
  {http://arxiv.org/abs/1702.04833} {arXiv:1702.04833 [gr-qc]} \BibitemShut
  {NoStop}%
\bibitem [{\citenamefont {Nakano}\ \emph {et~al.}(2017)\citenamefont {Nakano},
  \citenamefont {Sago}, \citenamefont {Tagoshi},\ and\ \citenamefont
  {Tanaka}}]{Nakano:2017fvh}%
  \BibitemOpen
  \bibfield  {author} {\bibinfo {author} {\bibfnamefont {H.}~\bibnamefont
  {Nakano}}, \bibinfo {author} {\bibfnamefont {N.}~\bibnamefont {Sago}},
  \bibinfo {author} {\bibfnamefont {H.}~\bibnamefont {Tagoshi}}, \ and\
  \bibinfo {author} {\bibfnamefont {T.}~\bibnamefont {Tanaka}},\ }\href
  {\doibase 10.1093/ptep/ptx093} {\bibfield  {journal} {\bibinfo  {journal}
  {PTEP}\ }\textbf {\bibinfo {volume} {2017}},\ \bibinfo {pages} {071E01}
  (\bibinfo {year} {2017})},\ \Eprint {http://arxiv.org/abs/1704.07175}
  {arXiv:1704.07175 [gr-qc]} \BibitemShut {NoStop}%
\bibitem [{\citenamefont {V{\"o}lkel}\ and\ \citenamefont
  {Kokkotas}(2017)}]{Volkel:2017kfj}%
  \BibitemOpen
  \bibfield  {author} {\bibinfo {author} {\bibfnamefont {S.~H.}\ \bibnamefont
  {V{\"o}lkel}}\ and\ \bibinfo {author} {\bibfnamefont {K.~D.}\ \bibnamefont
  {Kokkotas}},\ }\href {\doibase 10.1088/1361-6382/aa82de} {\bibfield
  {journal} {\bibinfo  {journal} {Class. Quant. Grav.}\ }\textbf {\bibinfo
  {volume} {34}},\ \bibinfo {pages} {175015} (\bibinfo {year} {2017})},\
  \Eprint {http://arxiv.org/abs/1704.07517} {arXiv:1704.07517 [gr-qc]}
  \BibitemShut {NoStop}%
\bibitem [{\citenamefont {Maselli}\ \emph {et~al.}(2017)\citenamefont
  {Maselli}, \citenamefont {V{\"o}lkel},\ and\ \citenamefont
  {Kokkotas}}]{Maselli:2017tfq}%
  \BibitemOpen
  \bibfield  {author} {\bibinfo {author} {\bibfnamefont {A.}~\bibnamefont
  {Maselli}}, \bibinfo {author} {\bibfnamefont {S.~H.}\ \bibnamefont
  {V{\"o}lkel}}, \ and\ \bibinfo {author} {\bibfnamefont {K.~D.}\ \bibnamefont
  {Kokkotas}},\ }\href {\doibase 10.1103/PhysRevD.96.064045} {\bibfield
  {journal} {\bibinfo  {journal} {Phys. Rev.}\ }\textbf {\bibinfo {volume}
  {D96}},\ \bibinfo {pages} {064045} (\bibinfo {year} {2017})},\ \Eprint
  {http://arxiv.org/abs/1708.02217} {arXiv:1708.02217 [gr-qc]} \BibitemShut
  {NoStop}%
\bibitem [{\citenamefont {Mark}\ \emph {et~al.}(2017)\citenamefont {Mark},
  \citenamefont {Zimmerman}, \citenamefont {Du},\ and\ \citenamefont
  {Chen}}]{Mark:2017dnq}%
  \BibitemOpen
  \bibfield  {author} {\bibinfo {author} {\bibfnamefont {Z.}~\bibnamefont
  {Mark}}, \bibinfo {author} {\bibfnamefont {A.}~\bibnamefont {Zimmerman}},
  \bibinfo {author} {\bibfnamefont {S.~M.}\ \bibnamefont {Du}}, \ and\ \bibinfo
  {author} {\bibfnamefont {Y.}~\bibnamefont {Chen}},\ }\href@noop {} {\
  (\bibinfo {year} {2017})},\ \Eprint {http://arxiv.org/abs/1706.06155}
  {arXiv:1706.06155 [gr-qc]} \BibitemShut {NoStop}%
\bibitem [{\citenamefont {Cardoso}\ and\ \citenamefont
  {Pani}(2017{\natexlab{a}})}]{Cardoso:2017njb}%
  \BibitemOpen
  \bibfield  {author} {\bibinfo {author} {\bibfnamefont {V.}~\bibnamefont
  {Cardoso}}\ and\ \bibinfo {author} {\bibfnamefont {P.}~\bibnamefont {Pani}},\
  }\href@noop {} {\  (\bibinfo {year} {2017}{\natexlab{a}})},\ \Eprint
  {http://arxiv.org/abs/1707.03021} {arXiv:1707.03021 [gr-qc]} \BibitemShut
  {NoStop}%
\bibitem [{\citenamefont {Cardoso}\ and\ \citenamefont
  {Pani}(2017{\natexlab{b}})}]{Cardoso:2017cqb}%
  \BibitemOpen
  \bibfield  {author} {\bibinfo {author} {\bibfnamefont {V.}~\bibnamefont
  {Cardoso}}\ and\ \bibinfo {author} {\bibfnamefont {P.}~\bibnamefont {Pani}},\
  }\href {\doibase 10.1038/s41550-017-0225-y} {\bibfield  {journal} {\bibinfo
  {journal} {Nat. Astron.}\ }\textbf {\bibinfo {volume} {1}},\ \bibinfo {pages}
  {586} (\bibinfo {year} {2017}{\natexlab{b}})},\ \Eprint
  {http://arxiv.org/abs/1709.01525} {arXiv:1709.01525 [gr-qc]} \BibitemShut
  {NoStop}%
\bibitem [{\citenamefont {Teukolsky}(1973)}]{Teukolsky:1973ha}%
  \BibitemOpen
  \bibfield  {author} {\bibinfo {author} {\bibfnamefont {S.~A.}\ \bibnamefont
  {Teukolsky}},\ }\href {\doibase 10.1086/152444} {\bibfield  {journal}
  {\bibinfo  {journal} {Astrophys. J.}\ }\textbf {\bibinfo {volume} {185}},\
  \bibinfo {pages} {635} (\bibinfo {year} {1973})}\BibitemShut {NoStop}%
\bibitem [{\citenamefont {Berti}\ \emph {et~al.}(2006)\citenamefont {Berti},
  \citenamefont {Cardoso},\ and\ \citenamefont {Casals}}]{Berti:2005gp}%
  \BibitemOpen
  \bibfield  {author} {\bibinfo {author} {\bibfnamefont {E.}~\bibnamefont
  {Berti}}, \bibinfo {author} {\bibfnamefont {V.}~\bibnamefont {Cardoso}}, \
  and\ \bibinfo {author} {\bibfnamefont {M.}~\bibnamefont {Casals}},\ }\href
  {\doibase 10.1103/PhysRevD.73.109902, 10.1103/PhysRevD.73.024013} {\bibfield
  {journal} {\bibinfo  {journal} {Phys. Rev.}\ }\textbf {\bibinfo {volume}
  {D73}},\ \bibinfo {pages} {024013} (\bibinfo {year} {2006})},\ \bibinfo
  {note} {[Erratum: Phys. Rev.D73,109902(2006)]},\ \Eprint
  {http://arxiv.org/abs/gr-qc/0511111} {arXiv:gr-qc/0511111 [gr-qc]}
  \BibitemShut {NoStop}%
\bibitem [{\citenamefont {Fishbach}\ \emph {et~al.}(2017)\citenamefont
  {Fishbach}, \citenamefont {Holz},\ and\ \citenamefont
  {Farr}}]{Fishbach:2017dwv}%
  \BibitemOpen
  \bibfield  {author} {\bibinfo {author} {\bibfnamefont {M.}~\bibnamefont
  {Fishbach}}, \bibinfo {author} {\bibfnamefont {D.~E.}\ \bibnamefont {Holz}},
  \ and\ \bibinfo {author} {\bibfnamefont {B.}~\bibnamefont {Farr}},\ }\href
  {\doibase 10.3847/2041-8213/aa7045} {\bibfield  {journal} {\bibinfo
  {journal} {Astrophys. J.}\ }\textbf {\bibinfo {volume} {840}},\ \bibinfo
  {pages} {L24} (\bibinfo {year} {2017})},\ \Eprint
  {http://arxiv.org/abs/1703.06869} {arXiv:1703.06869 [astro-ph.HE]}
  \BibitemShut {NoStop}%
\bibitem [{\citenamefont {Berti}\ \emph {et~al.}(2009)\citenamefont {Berti},
  \citenamefont {Cardoso},\ and\ \citenamefont {Starinets}}]{Berti:2009kk}%
  \BibitemOpen
  \bibfield  {author} {\bibinfo {author} {\bibfnamefont {E.}~\bibnamefont
  {Berti}}, \bibinfo {author} {\bibfnamefont {V.}~\bibnamefont {Cardoso}}, \
  and\ \bibinfo {author} {\bibfnamefont {A.~O.}\ \bibnamefont {Starinets}},\
  }\href {\doibase 10.1088/0264-9381/26/16/163001} {\bibfield  {journal}
  {\bibinfo  {journal} {Class. Quant. Grav.}\ }\textbf {\bibinfo {volume}
  {26}},\ \bibinfo {pages} {163001} (\bibinfo {year} {2009})},\ \Eprint
  {http://arxiv.org/abs/0905.2975} {arXiv:0905.2975 [gr-qc]} \BibitemShut
  {NoStop}%
\bibitem [{\citenamefont {Brito}\ \emph {et~al.}(2015)\citenamefont {Brito},
  \citenamefont {Cardoso},\ and\ \citenamefont {Pani}}]{Brito:2015oca}%
  \BibitemOpen
  \bibfield  {author} {\bibinfo {author} {\bibfnamefont {R.}~\bibnamefont
  {Brito}}, \bibinfo {author} {\bibfnamefont {V.}~\bibnamefont {Cardoso}}, \
  and\ \bibinfo {author} {\bibfnamefont {P.}~\bibnamefont {Pani}},\ }\href
  {\doibase 10.1007/978-3-319-19000-6} {\bibfield  {journal} {\bibinfo
  {journal} {Lect. Notes Phys.}\ }\textbf {\bibinfo {volume} {906}},\ \bibinfo
  {pages} {pp.1} (\bibinfo {year} {2015})},\ \Eprint
  {http://arxiv.org/abs/1501.06570} {arXiv:1501.06570 [gr-qc]} \BibitemShut
  {NoStop}%
\bibitem [{\citenamefont {Friedman}(1978)}]{Friedman:1978wla}%
  \BibitemOpen
  \bibfield  {author} {\bibinfo {author} {\bibfnamefont {J.~L.}\ \bibnamefont
  {Friedman}},\ }\href {\doibase 10.1007/BF01202527} {\bibfield  {journal}
  {\bibinfo  {journal} {Commun. Math. Phys.}\ }\textbf {\bibinfo {volume}
  {62}},\ \bibinfo {pages} {247} (\bibinfo {year} {1978})}\BibitemShut
  {NoStop}%
\bibitem [{\citenamefont {Cardoso}\ \emph {et~al.}(2008)\citenamefont
  {Cardoso}, \citenamefont {Pani}, \citenamefont {Cadoni},\ and\ \citenamefont
  {Cavaglia}}]{Cardoso:2007az}%
  \BibitemOpen
  \bibfield  {author} {\bibinfo {author} {\bibfnamefont {V.}~\bibnamefont
  {Cardoso}}, \bibinfo {author} {\bibfnamefont {P.}~\bibnamefont {Pani}},
  \bibinfo {author} {\bibfnamefont {M.}~\bibnamefont {Cadoni}}, \ and\ \bibinfo
  {author} {\bibfnamefont {M.}~\bibnamefont {Cavaglia}},\ }\href {\doibase
  10.1103/PhysRevD.77.124044} {\bibfield  {journal} {\bibinfo  {journal} {Phys.
  Rev.}\ }\textbf {\bibinfo {volume} {D77}},\ \bibinfo {pages} {124044}
  (\bibinfo {year} {2008})},\ \Eprint {http://arxiv.org/abs/0709.0532}
  {arXiv:0709.0532 [gr-qc]} \BibitemShut {NoStop}%
\bibitem [{\citenamefont {Maggio}\ \emph {et~al.}(2017)\citenamefont {Maggio},
  \citenamefont {Pani},\ and\ \citenamefont {Ferrari}}]{Maggio:2017ivp}%
  \BibitemOpen
  \bibfield  {author} {\bibinfo {author} {\bibfnamefont {E.}~\bibnamefont
  {Maggio}}, \bibinfo {author} {\bibfnamefont {P.}~\bibnamefont {Pani}}, \ and\
  \bibinfo {author} {\bibfnamefont {V.}~\bibnamefont {Ferrari}},\ }\href@noop
  {} {\  (\bibinfo {year} {2017})},\ \Eprint {http://arxiv.org/abs/1703.03696}
  {arXiv:1703.03696 [gr-qc]} \BibitemShut {NoStop}%
\bibitem [{\citenamefont {Harko}\ \emph {et~al.}(2013)\citenamefont {Harko},
  \citenamefont {Lobo}, \citenamefont {Mak},\ and\ \citenamefont
  {Sushkov}}]{Harko:2013yb}%
  \BibitemOpen
  \bibfield  {author} {\bibinfo {author} {\bibfnamefont {T.}~\bibnamefont
  {Harko}}, \bibinfo {author} {\bibfnamefont {F.~S.~N.}\ \bibnamefont {Lobo}},
  \bibinfo {author} {\bibfnamefont {M.~K.}\ \bibnamefont {Mak}}, \ and\
  \bibinfo {author} {\bibfnamefont {S.~V.}\ \bibnamefont {Sushkov}},\ }\href
  {\doibase 10.1103/PhysRevD.87.067504} {\bibfield  {journal} {\bibinfo
  {journal} {Phys. Rev.}\ }\textbf {\bibinfo {volume} {D87}},\ \bibinfo {pages}
  {067504} (\bibinfo {year} {2013})},\ \Eprint {http://arxiv.org/abs/1301.6878}
  {arXiv:1301.6878 [gr-qc]} \BibitemShut {NoStop}%
\bibitem [{\citenamefont {Lobo}\ and\ \citenamefont
  {Oliveira}(2009)}]{Lobo:2009ip}%
  \BibitemOpen
  \bibfield  {author} {\bibinfo {author} {\bibfnamefont {F.~S.~N.}\
  \bibnamefont {Lobo}}\ and\ \bibinfo {author} {\bibfnamefont {M.~A.}\
  \bibnamefont {Oliveira}},\ }\href {\doibase 10.1103/PhysRevD.80.104012}
  {\bibfield  {journal} {\bibinfo  {journal} {Phys. Rev.}\ }\textbf {\bibinfo
  {volume} {D80}},\ \bibinfo {pages} {104012} (\bibinfo {year} {2009})},\
  \Eprint {http://arxiv.org/abs/0909.5539} {arXiv:0909.5539 [gr-qc]}
  \BibitemShut {NoStop}%
\bibitem [{\citenamefont {Ayon-Beato}\ \emph {et~al.}(2016)\citenamefont
  {Ayon-Beato}, \citenamefont {Canfora},\ and\ \citenamefont
  {Zanelli}}]{Ayon-Beato:2015eca}%
  \BibitemOpen
  \bibfield  {author} {\bibinfo {author} {\bibfnamefont {E.}~\bibnamefont
  {Ayon-Beato}}, \bibinfo {author} {\bibfnamefont {F.}~\bibnamefont {Canfora}},
  \ and\ \bibinfo {author} {\bibfnamefont {J.}~\bibnamefont {Zanelli}},\ }\href
  {\doibase 10.1016/j.physletb.2015.11.065} {\bibfield  {journal} {\bibinfo
  {journal} {Phys. Lett.}\ }\textbf {\bibinfo {volume} {B752}},\ \bibinfo
  {pages} {201} (\bibinfo {year} {2016})},\ \Eprint
  {http://arxiv.org/abs/1509.02659} {arXiv:1509.02659 [gr-qc]} \BibitemShut
  {NoStop}%
\bibitem [{\citenamefont {Canfora}\ \emph {et~al.}(2017)\citenamefont
  {Canfora}, \citenamefont {Dimakis},\ and\ \citenamefont
  {Paliathanasis}}]{Canfora:2017gno}%
  \BibitemOpen
  \bibfield  {author} {\bibinfo {author} {\bibfnamefont {F.}~\bibnamefont
  {Canfora}}, \bibinfo {author} {\bibfnamefont {N.}~\bibnamefont {Dimakis}}, \
  and\ \bibinfo {author} {\bibfnamefont {A.}~\bibnamefont {Paliathanasis}},\
  }\href {\doibase 10.1103/PhysRevD.96.025021} {\bibfield  {journal} {\bibinfo
  {journal} {Phys. Rev.}\ }\textbf {\bibinfo {volume} {D96}},\ \bibinfo {pages}
  {025021} (\bibinfo {year} {2017})},\ \Eprint
  {http://arxiv.org/abs/1707.02270} {arXiv:1707.02270 [hep-th]} \BibitemShut
  {NoStop}%
\bibitem [{\citenamefont {Berti}\ and\ \citenamefont
  {Cardoso}(2006)}]{Berti:2006qt}%
  \BibitemOpen
  \bibfield  {author} {\bibinfo {author} {\bibfnamefont {E.}~\bibnamefont
  {Berti}}\ and\ \bibinfo {author} {\bibfnamefont {V.}~\bibnamefont
  {Cardoso}},\ }\href {\doibase 10.1142/S0218271806009637} {\bibfield
  {journal} {\bibinfo  {journal} {Int. J. Mod. Phys.}\ }\textbf {\bibinfo
  {volume} {D15}},\ \bibinfo {pages} {2209} (\bibinfo {year} {2006})},\ \Eprint
  {http://arxiv.org/abs/gr-qc/0605101} {arXiv:gr-qc/0605101 [gr-qc]}
  \BibitemShut {NoStop}%
\bibitem [{\citenamefont {Macedo}\ \emph
  {et~al.}(2013{\natexlab{a}})\citenamefont {Macedo}, \citenamefont {Pani},
  \citenamefont {Cardoso},\ and\ \citenamefont {Crispino}}]{Macedo:2013qea}%
  \BibitemOpen
  \bibfield  {author} {\bibinfo {author} {\bibfnamefont {C.~F.~B.}\
  \bibnamefont {Macedo}}, \bibinfo {author} {\bibfnamefont {P.}~\bibnamefont
  {Pani}}, \bibinfo {author} {\bibfnamefont {V.}~\bibnamefont {Cardoso}}, \
  and\ \bibinfo {author} {\bibfnamefont {L.~C.~B.}\ \bibnamefont {Crispino}},\
  }\href {\doibase 10.1088/0004-637X/774/1/48} {\bibfield  {journal} {\bibinfo
  {journal} {Astrophys. J.}\ }\textbf {\bibinfo {volume} {774}},\ \bibinfo
  {pages} {48} (\bibinfo {year} {2013}{\natexlab{a}})},\ \Eprint
  {http://arxiv.org/abs/1302.2646} {arXiv:1302.2646 [gr-qc]} \BibitemShut
  {NoStop}%
\bibitem [{\citenamefont {Macedo}\ \emph
  {et~al.}(2013{\natexlab{b}})\citenamefont {Macedo}, \citenamefont {Pani},
  \citenamefont {Cardoso},\ and\ \citenamefont {Crispino}}]{Macedo:2013jja}%
  \BibitemOpen
  \bibfield  {author} {\bibinfo {author} {\bibfnamefont {C.~F.~B.}\
  \bibnamefont {Macedo}}, \bibinfo {author} {\bibfnamefont {P.}~\bibnamefont
  {Pani}}, \bibinfo {author} {\bibfnamefont {V.}~\bibnamefont {Cardoso}}, \
  and\ \bibinfo {author} {\bibfnamefont {L.~C.~B.}\ \bibnamefont {Crispino}},\
  }\href {\doibase 10.1103/PhysRevD.88.064046} {\bibfield  {journal} {\bibinfo
  {journal} {Phys. Rev.}\ }\textbf {\bibinfo {volume} {D88}},\ \bibinfo {pages}
  {064046} (\bibinfo {year} {2013}{\natexlab{b}})},\ \Eprint
  {http://arxiv.org/abs/1307.4812} {arXiv:1307.4812 [gr-qc]} \BibitemShut
  {NoStop}%
\bibitem [{\citenamefont {Chirenti}\ and\ \citenamefont
  {Rezzolla}(2016)}]{Chirenti:2016hzd}%
  \BibitemOpen
  \bibfield  {author} {\bibinfo {author} {\bibfnamefont {C.}~\bibnamefont
  {Chirenti}}\ and\ \bibinfo {author} {\bibfnamefont {L.}~\bibnamefont
  {Rezzolla}},\ }\href {\doibase 10.1103/PhysRevD.94.084016} {\bibfield
  {journal} {\bibinfo  {journal} {Phys. Rev.}\ }\textbf {\bibinfo {volume}
  {D94}},\ \bibinfo {pages} {084016} (\bibinfo {year} {2016})},\ \Eprint
  {http://arxiv.org/abs/1602.08759} {arXiv:1602.08759 [gr-qc]} \BibitemShut
  {NoStop}%
\bibitem [{\citenamefont {Konoplya}\ and\ \citenamefont
  {Zhidenko}(2016)}]{Konoplya:2016hmd}%
  \BibitemOpen
  \bibfield  {author} {\bibinfo {author} {\bibfnamefont {R.~A.}\ \bibnamefont
  {Konoplya}}\ and\ \bibinfo {author} {\bibfnamefont {A.}~\bibnamefont
  {Zhidenko}},\ }\href {\doibase 10.1088/1475-7516/2016/12/043} {\bibfield
  {journal} {\bibinfo  {journal} {JCAP}\ }\textbf {\bibinfo {volume} {1612}},\
  \bibinfo {pages} {043} (\bibinfo {year} {2016})},\ \Eprint
  {http://arxiv.org/abs/1606.00517} {arXiv:1606.00517 [gr-qc]} \BibitemShut
  {NoStop}%
\bibitem [{\citenamefont {Nandi}\ \emph {et~al.}(2017)\citenamefont {Nandi},
  \citenamefont {Izmailov}, \citenamefont {Yanbekov},\ and\ \citenamefont
  {Shayakhmetov}}]{Nandi:2016uzg}%
  \BibitemOpen
  \bibfield  {author} {\bibinfo {author} {\bibfnamefont {K.~K.}\ \bibnamefont
  {Nandi}}, \bibinfo {author} {\bibfnamefont {R.~N.}\ \bibnamefont {Izmailov}},
  \bibinfo {author} {\bibfnamefont {A.~A.}\ \bibnamefont {Yanbekov}}, \ and\
  \bibinfo {author} {\bibfnamefont {A.~A.}\ \bibnamefont {Shayakhmetov}},\
  }\href {\doibase 10.1103/PhysRevD.95.104011} {\bibfield  {journal} {\bibinfo
  {journal} {Phys. Rev.}\ }\textbf {\bibinfo {volume} {D95}},\ \bibinfo {pages}
  {104011} (\bibinfo {year} {2017})},\ \Eprint
  {http://arxiv.org/abs/1611.03479} {arXiv:1611.03479 [gr-qc]} \BibitemShut
  {NoStop}%
\bibitem [{\citenamefont {Barcel{\'o}}\ \emph {et~al.}(2017)\citenamefont
  {Barcel{\'o}}, \citenamefont {Carballo-Rubio},\ and\ \citenamefont
  {Garay}}]{Barcelo:2017lnx}%
  \BibitemOpen
  \bibfield  {author} {\bibinfo {author} {\bibfnamefont {C.}~\bibnamefont
  {Barcel{\'o}}}, \bibinfo {author} {\bibfnamefont {R.}~\bibnamefont
  {Carballo-Rubio}}, \ and\ \bibinfo {author} {\bibfnamefont {L.~J.}\
  \bibnamefont {Garay}},\ }\href {\doibase 10.1007/JHEP05(2017)054} {\bibfield
  {journal} {\bibinfo  {journal} {JHEP}\ }\textbf {\bibinfo {volume} {05}},\
  \bibinfo {pages} {054} (\bibinfo {year} {2017})},\ \Eprint
  {http://arxiv.org/abs/1701.09156} {arXiv:1701.09156 [gr-qc]} \BibitemShut
  {NoStop}%
\bibitem [{\citenamefont {Brustein}\ \emph {et~al.}(2017)\citenamefont
  {Brustein}, \citenamefont {Medved},\ and\ \citenamefont
  {Yagi}}]{Brustein:2017koc}%
  \BibitemOpen
  \bibfield  {author} {\bibinfo {author} {\bibfnamefont {R.}~\bibnamefont
  {Brustein}}, \bibinfo {author} {\bibfnamefont {A.~J.~M.}\ \bibnamefont
  {Medved}}, \ and\ \bibinfo {author} {\bibfnamefont {K.}~\bibnamefont
  {Yagi}},\ }\href {\doibase 10.1103/PhysRevD.96.064033} {\bibfield  {journal}
  {\bibinfo  {journal} {Phys. Rev.}\ }\textbf {\bibinfo {volume} {D96}},\
  \bibinfo {pages} {064033} (\bibinfo {year} {2017})},\ \Eprint
  {http://arxiv.org/abs/1704.05789} {arXiv:1704.05789 [gr-qc]} \BibitemShut
  {NoStop}%
\bibitem [{\citenamefont {Bena}\ and\ \citenamefont
  {Warner}(2008)}]{Bena:2007kg}%
  \BibitemOpen
  \bibfield  {author} {\bibinfo {author} {\bibfnamefont {I.}~\bibnamefont
  {Bena}}\ and\ \bibinfo {author} {\bibfnamefont {N.~P.}\ \bibnamefont
  {Warner}},\ }\bibfield  {booktitle} {\emph {\bibinfo {booktitle} {{Winter
  School on Attractor Mechanism (SAM 2006) Frascati, Italy, March 20-24,
  2006}}},\ }\href {\doibase 10.1007/978-3-540-79523-0_1} {\bibfield  {journal}
  {\bibinfo  {journal} {Lect. Notes Phys.}\ }\textbf {\bibinfo {volume}
  {755}},\ \bibinfo {pages} {1} (\bibinfo {year} {2008})},\ \Eprint
  {http://arxiv.org/abs/hep-th/0701216} {arXiv:hep-th/0701216 [hep-th]}
  \BibitemShut {NoStop}%
\bibitem [{\citenamefont {Skenderis}\ and\ \citenamefont
  {Taylor}(2008)}]{Skenderis:2008qn}%
  \BibitemOpen
  \bibfield  {author} {\bibinfo {author} {\bibfnamefont {K.}~\bibnamefont
  {Skenderis}}\ and\ \bibinfo {author} {\bibfnamefont {M.}~\bibnamefont
  {Taylor}},\ }\href {\doibase 10.1016/j.physrep.2008.08.001} {\bibfield
  {journal} {\bibinfo  {journal} {Phys. Rept.}\ }\textbf {\bibinfo {volume}
  {467}},\ \bibinfo {pages} {117} (\bibinfo {year} {2008})},\ \Eprint
  {http://arxiv.org/abs/0804.0552} {arXiv:0804.0552 [hep-th]} \BibitemShut
  {NoStop}%
\bibitem [{\citenamefont {Mathur}(2008)}]{Mathur:2008nj}%
  \BibitemOpen
  \bibfield  {author} {\bibinfo {author} {\bibfnamefont {S.~D.}\ \bibnamefont
  {Mathur}},\ }\href@noop {} {\  (\bibinfo {year} {2008})},\ \Eprint
  {http://arxiv.org/abs/0810.4525} {arXiv:0810.4525 [hep-th]} \BibitemShut
  {NoStop}%
\bibitem [{\citenamefont {Balasubramanian}\ \emph {et~al.}(2008)\citenamefont
  {Balasubramanian}, \citenamefont {de~Boer}, \citenamefont {El-Showk},\ and\
  \citenamefont {Messamah}}]{Balasubramanian:2008da}%
  \BibitemOpen
  \bibfield  {author} {\bibinfo {author} {\bibfnamefont {V.}~\bibnamefont
  {Balasubramanian}}, \bibinfo {author} {\bibfnamefont {J.}~\bibnamefont
  {de~Boer}}, \bibinfo {author} {\bibfnamefont {S.}~\bibnamefont {El-Showk}}, \
  and\ \bibinfo {author} {\bibfnamefont {I.}~\bibnamefont {Messamah}},\
  }\bibfield  {booktitle} {\emph {\bibinfo {booktitle} {{Strings, supergravity
  and gauge theories. Proceedings, European RTN Winter School, CERN, Geneva,
  Switzerland, January 21-25, 2008}}},\ }\href {\doibase
  10.1088/0264-9381/25/21/214004} {\bibfield  {journal} {\bibinfo  {journal}
  {Class. Quant. Grav.}\ }\textbf {\bibinfo {volume} {25}},\ \bibinfo {pages}
  {214004} (\bibinfo {year} {2008})},\ \Eprint {http://arxiv.org/abs/0811.0263}
  {arXiv:0811.0263 [hep-th]} \BibitemShut {NoStop}%
\bibitem [{\citenamefont {Bena}\ \emph {et~al.}(2017)\citenamefont {Bena},
  \citenamefont {Turton}, \citenamefont {Walker},\ and\ \citenamefont
  {Warner}}]{Bena:2017upb}%
  \BibitemOpen
  \bibfield  {author} {\bibinfo {author} {\bibfnamefont {I.}~\bibnamefont
  {Bena}}, \bibinfo {author} {\bibfnamefont {D.}~\bibnamefont {Turton}},
  \bibinfo {author} {\bibfnamefont {R.}~\bibnamefont {Walker}}, \ and\ \bibinfo
  {author} {\bibfnamefont {N.~P.}\ \bibnamefont {Warner}},\ }\href@noop {} {\
  (\bibinfo {year} {2017})},\ \Eprint {http://arxiv.org/abs/1709.01107}
  {arXiv:1709.01107 [hep-th]} \BibitemShut {NoStop}%
\bibitem [{\citenamefont {Hertog}\ and\ \citenamefont
  {Hartle}(2017)}]{Hertog:2017vod}%
  \BibitemOpen
  \bibfield  {author} {\bibinfo {author} {\bibfnamefont {T.}~\bibnamefont
  {Hertog}}\ and\ \bibinfo {author} {\bibfnamefont {J.}~\bibnamefont
  {Hartle}},\ }\href@noop {} {\  (\bibinfo {year} {2017})},\ \Eprint
  {http://arxiv.org/abs/1704.02123} {arXiv:1704.02123 [hep-th]} \BibitemShut
  {NoStop}%
\bibitem [{\citenamefont {Hertog}\ \emph {et~al.}(2017)\citenamefont {Hertog},
  \citenamefont {Lemmens},\ and\ \citenamefont {Vercnocke}}]{Hertog_upcoming}%
  \BibitemOpen
  \bibfield  {author} {\bibinfo {author} {\bibfnamefont {T.}~\bibnamefont
  {Hertog}}, \bibinfo {author} {\bibfnamefont {T.}~\bibnamefont {Lemmens}}, \
  and\ \bibinfo {author} {\bibfnamefont {B.}~\bibnamefont {Vercnocke}},\
  }\bibfield  {booktitle} {\emph {\bibinfo {booktitle} {{in preparation}}},\
  }\href@noop {} {\  (\bibinfo {year} {2017})}\BibitemShut {NoStop}%
\bibitem [{\citenamefont {Poschl}\ and\ \citenamefont
  {Teller}(1933)}]{Poschl:1933zz}%
  \BibitemOpen
  \bibfield  {author} {\bibinfo {author} {\bibfnamefont {G.}~\bibnamefont
  {Poschl}}\ and\ \bibinfo {author} {\bibfnamefont {E.}~\bibnamefont
  {Teller}},\ }\href {\doibase 10.1007/BF01331132} {\bibfield  {journal}
  {\bibinfo  {journal} {Z. Phys.}\ }\textbf {\bibinfo {volume} {83}},\ \bibinfo
  {pages} {143} (\bibinfo {year} {1933})}\BibitemShut {NoStop}%
\end{thebibliography}%
\label{biblio}

\end{document}